\documentclass[fleqn,usenatbib]{mnras}

\usepackage{newtxtext,newtxmath}

\usepackage[T1]{fontenc}
\usepackage{ae,aecompl}

\usepackage{graphicx}	
\usepackage{amsmath}	
\usepackage{bm}	        
\usepackage{amssymb}	
\usepackage{multirow}
\usepackage[dvipsnames]{xcolor}
\usepackage{hyperref}

\newcommand{\kb}{k_{\text{B}}}
\newcommand{\D}{\text{d}}
\newcommand{\x}{\textbf{x}}
\newcommand{\n}{\hat{\textbf{n}}}

\newcommand{\Magritte}{{\sc Magritte}}

\newcommand{\orcid}[1]{{\hskip.5mm \href{#1}{\includegraphics[height=8px]{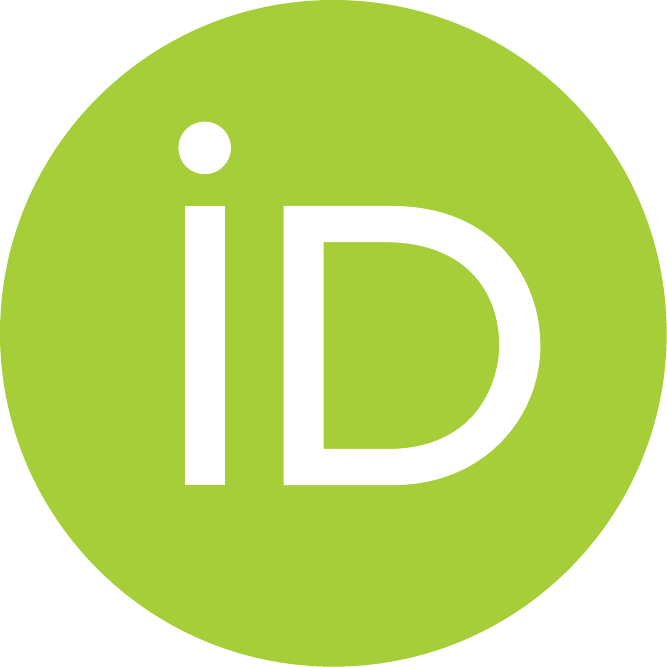}} \hskip.5mm}}

\title
[\textsc{Magritte}: NLTE atomic and molecular line modelling]
{\textsc{Magritte}, a modern software library for 3D radiative transfer: \\
I. Non-LTE atomic and molecular line modelling}

\author
[F. De Ceuster et al.]
{Frederik De Ceuster$^{\orcid{https://orcid.org/0000-0001-5887-8498} 1, 2}$\thanks{Contact e-mail: \href{frederik.deceuster@kuleuven.be}{frederik.deceuster@kuleuven.be}},
 Ward Homan$^{\orcid{https://orcid.org/0000-0001-7314-5081} 2}$,
 Jeremy Yates$^{\orcid{https://orcid.org/0000-0003-1954-8749} 1}$,
 Leen Decin$^{\orcid{https://orcid.org/0000-0002-5342-8612} 2, 3}$, \newauthor
 Peter Boyle$^{\orcid{https://orcid.org/0000-0002-8960-1587} 4}$, and
 James Hetherington$^{\orcid{https://orcid.org/0000-0001-6993-0319} 5, 6}$
  \\ \\
  $^{1}$ Department of Physics and Astronomy, University College London, Gower Place, London, WC1E 6BT, UK \\
  $^{2}$ Department of Physics and Astronomy, Institute of Astronomy, KU Leuven, Celestijnenlaan 200D, 3001 Leuven, Belgium \\
  $^{3}$ School of Chemistry, University of Leeds, Leeds LS2 9JT, UK \\
  $^{4}$ School of Physics and Astronomy, The University of Edinburgh, Edinburgh EH9 3FD, UK \\
  $^{5}$ Department of Computer Science, University College London, Bloomsburry, London, WC1E 6EA, UK \\
  $^{6}$ The Alan Turing Institute, 96 Euston Road, Kings Cross, London, NW1 2DB, UK
}

\date{Accepted XXX. Received YYY; in original form ZZZ}

\pubyear{2019}

\begin{document}
\label{firstpage}
\pagerange{\pageref{firstpage}--\pageref{lastpage}}
\maketitle

\begin{abstract}
    Radiative transfer is a key component in almost all astrophysical and cosmological simulations. We present \Magritte{}: a modern open-source software library for 3D radiative transfer. It uses a deterministic ray-tracer and formal solver, i.e. it computes the radiation field by tracing rays through the model and solving the radiative transfer equation in its second-order form along a fixed set of rays originating from each point. \Magritte{} can handle structured and unstructured input meshes, as well as smoothed-particle hydrodynamics (SPH) particle data. In this first paper, we describe the numerical implementation, semi-analytic tests and cross-code benchmarks for the non-LTE line radiative transfer module of \Magritte{}. This module uses the radiative transfer solver to self-consistently determine the populations of the quantised energy levels of atoms and molecules using an accelerated Lambda iteration (ALI) scheme. We compare \Magritte{} with the established radiative transfer solvers \textsc{Ratran} (1D) and \textsc{Lime} (3D) on the van Zadelhoff benchmark and present a first application to a simple Keplerian disc model. Comparing with \textsc{Lime}, we conclude that \Magritte{} produces more accurate and more precise results, especially at high optical depth, and that it is faster.
\end{abstract}




\section{Introduction}

Radiative processes play an essential role in the dynamics, chemistry and energy balance of various astrophysical objects, from planetary and stellar atmospheres to galaxies and the Universe as a whole.
Radiation can provide a radiative pressure that can drive dynamics \citep[see e.g.][]{Hoefner2003}, it can affect chemistry through various photo-ionisation and photo-dissociation reactions \citep[see e.g.][]{Huggins1982}, and it can efficiently heat or cool very localised regions \cite[see e.g.][]{Woitke1996}.
Furthermore, the radiative properties determine what can and cannot be seen in observations, i.e. which regions are visible in what part of the electromagnetic spectrum.

Ever since the first detection and identification of atoms and molecules in space \citep{Douglas1941, Weinreb1963}, their line emission and absorption features have been an indispensable diagnostic tool to infer the physical and chemical conditions throughout the Universe.
In order to interpret the observational data, we require high precision atomic and molecular data \citep{Schoier2005} as well as sound theoretical models.
Historically, these models quickly evolved from highly idealised equilibrium systems to more self-consistent non-equilibrium models \citep{Mihalas1973}.
Moreover, with the advent of high (spatial) resolution imaging, for instance using the Atacama Large Millimetre Array (ALMA), full 3D models are imperative to properly model the intricate structures observed in the data \citep[see e.g.][]{Alves2019, Smith2019, Decin2015, Maercker2012}.

Given the tight coupling between radiation field and medium, it is crucial both in astrophysical and cosmological modelling to properly account for all radiative processes and their interdependence.
This, however, can be highly complicated due to: i) an intricate 3D geometrical structure shielding or exposing specific regions to radiation, ii) the scattering of radiation by dust or free electrons yielding additional non-trivial coupling between the geometry and the radiation field, and iii) the mixing in frequency space due to Doppler shifts caused by velocity gradients in the medium.
Furthermore, the coupling between the radiative processes and the often very specialised and diverse dynamical and chemical models requires a modular radiative transfer solver that can easily be integrated with the various existing hydrodynamics and chemistry models.
Finally, the ever growing size and complexity of these models requires fast and scalable algorithms that can efficiently leverage the wealth of modern computational resources.


There are two main computational strategies to solve radiative transfer problems.
On the one hand there are probabilistic (Monte Carlo) solvers such as e.g. \textsc{RadMC-3D}   \citep{Dullemond2012},
\textsc{Skirt}      \citep{Verstocken2017},
\textsc{CMacIonize} \citep{Vandenbroucke2018}, and some components of
\textsc{Torus}      \citep{Harries2019}.
On the other hand there are deterministic or formal solvers such as e.g. \textsc{SPHray}     \citep{Altay2008},
\textsc{3D-pdr}     \citep{Bisbas2012}, and
\textsc{Lampray}    \citep{Frostholm2018}.
Furthermore, there are also codes that combine ideas from both techniques such as \textsc{Ratran} \citep{Hogerheijde2000} and its 3D successor \textsc{Lime} \citep{Brinch2010}.
The latter has been widely used to model atomic and molecular lines in 3D models of various astrophysical objects
\citep[see e.g.][]{
Booth2019,
Montarges2019,
Homan2018,
Evans2018,
Walsh2016,
Bergin2013,
Maercker2012,
Andrews2012}.

Currently, most radiative transfer solvers use probabilistic methods.
These methods mimic the physical photon transport by propagating a number of photon packets through the medium \citep[see e.g.][for an extensive review]{Noebauer2019}.
The main issue with this approach is that the trajectories of these photon packets are randomly determined by the properties of the medium.
This implies that they can get trapped in opaque regions, impeding them from contributing much to the overall radiation field.
Hence, a large number of packets need to be propagated which can significantly increase the computation time.
Although many techniques have been devised to avoid the trapping of photon packets \citep[see e.g.][]{Yusef1984}, it remains challenging for probabilistic radiative transfer solvers to efficiently obtain accurate results, especially at medium to high optical depths \citep{Camps2018}.

Deterministic or formal solvers compute the radiation field by solving the radiative transfer equation along rays through the medium. Since the optical properties of the medium often depend on the radiation field this has to be done in an iterative way. Although there are no photon packets in this approach, a problem physically very similar to photon trapping can manifest itself in the form of slow convergence of the iteration process.
In this context, the problem was first identified in the 1970s by various authors \citep[see e.g.][and references therein]{Scharmer1985} and is more commonly known as the Lamba-iteration problem.
This problem arose when attempts were made to model the radiative hydrodynamics of hot stars without assuming local thermodynamic equilibrium (i.e. non-LTE).
The extremely slow or false convergence produced by this effect resulted in erroneous fits to the observed data.
Subsequent work, for instance by \cite{Olson1986} and \cite{Rybicki1991}, elegantly addressed these issues using a technique called accelerated Lambda iteration (ALI). For a complete overview of these methods see, for example, \cite{Hubeny2014}.

Although deterministic solvers could better cope with the optical depth related issues, probabilistic solvers became more popular due to their relative ease of implementation, especially in two and three spatial dimensions.
However, with the development of fast solution methods it is now possible to implement a deterministic solver with comparative ease.
When combined with the ability to sample rays finer, multi-dimensional ray-tracing codes can now become powerful probes of objects with complex geometries, velocity fields, and optical depth ranges.
Moreover, their deterministic computational scheme leads to various opportunities for optimisation and facilitates utilising the various layers of parallelism in the calculation, further reducing the computational cost.

\Magritte{} is a modern open-source software library for 3D radiative transfer. It is written in \textsc{C++11}, but almost all classes and functions are wrapped using \textsc{PyBind11} \citep{pybind11} such that they can also be used in \textsc{Python}.
Our motivation to develop \Magritte{} is twofold.
On the one hand, the ever increasing amount of high quality observational data puts increasingly higher demands on the modelling software, while, on the other hand, advances in computer technology provide us with the means to meet these demands. Common examples are the extended use of different layers of parallelism (e.g. vector instructions, multi-threading and message passing) and the growing availability of hardware accelerators such as graphics processing units (GPUs) or field programmable gate arrays (FPGAs).
Using these technologies in an existing code base, however, often requires a complete rewrite of the internal data structures.
Therefore, we opted to build a new code base that is flexible enough to cope with the requirements for multiple astrophysical and cosmological applications and has a modular data structure that can readily be adapted to leverage the different forms of parallelism and hardware accelerators available in modern (super)computer architectures.

Since advances in modelling are increasingly made by improved software implementations rather than new mathematical techniques, it is imperative that both the software and its source code are publicly available for the community to review and adapt.
Therefore, we commit ourselves to make future releases of \Magritte{} and its source code publicly available\footnote{\label{GNUgplv2}
Under \href{https://www.gnu.org/licenses/old-licenses/gpl-2.0.en.html}{GNU General Public License v2.0}.} at \href{https://github.com/Magritte-code}{github.com/Magritte-code}.

This is the first paper in a series in which we will analyse the physical, mathematical and computational aspects of the various components of the software library. In this first paper, we present \Magritte{}'s module for atomic and molecular line radiative transfer. The radiation field is computed self-consistently with the populations of the quantised energy levels. In contrast to many early treatments of line radiative transfer, this approach does not make the assumption of local thermodynamic equilibrium (LTE), which hence classifies it as a non-LTE solver. We present \Magritte{}'s ray-tracing scheme which only uses cell locations and nearest neighbour information. Therefore, it can easily cope with smoothed particle hydrodynamics (SPH) particles as well as structured and unstructured model meshes. We introduce our solution method to solve the radiative transfer equation along a ray pair and present our implementation of the accelerated Lambda iteration (ALI) scheme based on \cite{Rybicki1991}. To validate our methods, we run a set of test models for which we can obtain semi-analytical results. This way, we can get an absolute measure of the errors resulting from our methods. We further demonstrate \Magritte{}'s validity, by performing a cross-code comparison with \textsc{Ratran} \citep{Hogerheijde2000} and \textsc{Lime} \citep{Brinch2010} using the \cite{vanZadelhoff2002} benchmark, and some additional variations on that. Finally, we present a first application of \Magritte{} modelling the CO emission of a simple Keplerian disc.


The structure of this paper is as follows.
In Section \ref{sec:physical_problem}, we introduce the radiative transfer problem and elaborate on the tight coupling between the radiation field and the medium.
Section \ref{sec:numerical_implementation} presents our solution methods to the problem and the numerical implementations.
In Section \ref{sec:tests_and_benchmarks}, we describe a set of semi-analytic tests and cross-code benchmarks to validate our methods and Section \ref{sec:application} describes a first application of \Magritte{} to a simple Keplerian disc model.
Finally, our results are discussed in Section \ref{sec:discussion} and we conclude with Section \ref{sec:conclusions}.

\section{Physical problem}
\label{sec:physical_problem}

\subsection{Radiative transfer}

The objective of radiative transfer is to determine the radiation field in a region, given the properties of the medium in that region and some boundary conditions.
The radiation field is described in terms of its specific monochromatic intensity $I_{\nu}(\x,\n)$, i.e. the energy transported in a certain direction in a certain frequency bin.
This is a function of frequency ($\nu$), position ($\x$) and direction ($\n$). Any interaction between the radiation field and the medium can be described in terms of the change to the specific monochromatic intensity.
The radiative transfer equation relates this change in specific monochromatic intensity $I_{\nu}(\x,\n)$ along a ray in direction $\n$ to the local emissivity $\eta_{\nu}(\x)$ and opacity $\chi_{\nu}(\x)$ of the medium.
Scattering introduces an extra contribution to both the emissivity and opacity \citep{Chandrasekhar1960,Steinacker2013}. The time-independent radiative transfer equation including scattering reads
\begin{equation}
  \begin{split}
    \n \cdot \nabla I_{\nu}(\n) \ = \ \ &\eta_{\nu} \ - \ \big( \chi_{\nu} \ + \ \chi^{\text{sca}}_{\nu}(\n) \big) \ I_{\nu}(\n) \\
    &+ \  \oint \D\Omega' \int_{0}^{\infty} \D \nu' \ \Phi_{\nu \, \nu'}(\n \, ,\n') \ I_{\nu'}(\n') ,
  \end{split}
  \label{eq:RTE}
\end{equation}
where $\chi^{\text{sca}}_{\nu}(\n)$ is the extra opacity due to scattering and $\Phi_{\nu \, \nu'}(\n \, ,\n')$ is the scattering redistribution function which gives the probability for radiation of frequency $\nu'$ incoming along direction $\n'$ to be scattered in direction $\n$ and to be shifted to frequency $\nu$.

For local radiative processes, we assume both the emissivity and opacity to be isotropic, i.e. independent of the direction $\n$. However, in contrast to the classical general formulation of the transfer equation by \cite{Cannon1971, Cannon1972}, we allowed for a directional dependence in the scattering opacity. This more general approach allows us to also treat, for instance, scattering from dust grains that are aligned by a magnetic field \citep[see e.g.][and the references there]{Andersson2015}. The anisotropy of the scattering opacity slightly complicates the solution methods. However, we have included it to keep our solution methods as general as possible.

The radiative transfer equation (\ref{eq:RTE}) is a first-order integro-differential equation. Generally, it can only be solved in an iterative way, since both the emissivity and opacity depend on the radiation field. We discuss our solution strategy for solving the transfer equation in section \ref{subsec:solving_the_transfer_equation}. First, we break down the coupling between the medium and the radiation field.

\subsection{Coupling radiation field \& medium}

In general we can distinguish four types of interactions between a radiation field and a medium based on the frequency range on which they act: line, ionisation, continuum and scattering interactions. In this first paper, we will limit ourselves to atomic and molecular line interactions.

\subsubsection{Atomic and molecular lines}

Electronic, rotational and vibrational transitions between the quantized energy levels in atomic and molecular species can lead to significant emission and absorption in narrow frequency ranges. These transitions are referred to as line transitions, due to the characteristically narrow features they induce in spectra.

The resulting emissivity and opacity due to a line transition from a level $i$ to a level $j$ (with level energies $E_{i}>E_{j}$) are given in terms of the Einstein $A_{ij}$, $B_{ji}$ and $B_{ij}$ coefficients and the populations $n_{i}(\x)$ of the quantized energy levels
\begin{equation}
\begin{split}
    \eta^{ij}_{\nu}(\x) \ &= \ \frac{h\nu}{4\pi} \  n_{i}(\x) \ A_{ij} \ \phi^{ij}_{\nu}(\x), \\
    \chi^{ij}_{\nu}(\x) \ &= \ \frac{h\nu}{4\pi} \ \left(n_{j}(\x) \ B_{ji} \ - \ n_{i}(\x) \ B_{ij}\right) \phi^{ij}_{\nu}(\x).
\end{split}
\label{eq:lines}
\end{equation}
where $A_{ij}$ and $B_{ij}$ account, respectively, for spontaneous and stimulated emission and $B_{ji}$ accounts for absorption. Note that stimulated emission is treated as negative absorption. Both line emissivity and opacity are proportional to the line profile function $\phi^{ij}_{\nu}(\x)$. In this paper, we assume Gaussian line profile functions\footnote{However, all our methods can also readily be applied to all other types of line profile functions.} resulting from the Doppler shifts caused by the thermal and turbulent motions of the atoms and molecules in the medium,
\begin{equation}
  \phi_{\nu}^{ij}(\x) \ = \ \frac{1}{\delta \nu_{ij}(\x)\sqrt{\pi}} \ \exp \left[-\left(\frac{\nu-\nu_{ij}} {\delta\nu_{ij}(\x)}\right)^{2}\right],
\end{equation}
where the characteristic line width
\begin{equation}
    \delta\nu_{ij}(\x) \ = \ \frac{\nu_{ij}}{c} \ \sqrt{ \varv_{\text{thermal}}(\x)^{2} \ + \ \varv_{\text{turbulent}}(\x)^{2}},
\end{equation}
is determined by the mean thermal and turbulent velocities of the medium in the co-moving frame.

Many early line radiative transfer models assumed the populations of the quantized energy levels to be in local thermodynamic equilibrium (LTE), i.e. particle velocities, level populations, and radiation field are completely determined by the local gas temperature. In contrast, we will only assume kinetic equilibrium, i.e. we only assume a Maxwell-Boltzmann distribution for the particle velocities. This situation is often referred to as non-LTE. As a result, the mean local velocity of the gas particles in the co-moving frame can be characterised by
\begin{equation}
    \varv_{\text{thermal}}(\x) = \sqrt{\frac{2 \kb T(\x)}{ m_{\text{spec}}}},
\end{equation}
where $m_{\text{spec}}$ is the mass of the species of gas under consideration.
If we make no further assumptions on the level populations, they can only be determined by directly solving the kinetic rate equations, which, in the co-moving frame, are given by,
\begin{equation}
	\frac{\partial n_{i}(\x)}{\partial t} \ = \
	\sum_{j=1}^{N} n_{j}(\x) P_{ji}(\x) \ - \ n_{i}(\x) \sum_{j=1}^{N} P_{ij}(\x).
\label{eq:levelpop}
\end{equation}
The components of the matrix $ P_{ij}(\x)$ denote the transition rates from level $i$ to level $j$. Hence, for each level $i$, $P_{ii}(\x) = 0$. The transition rates are composed of a radiative part $R_{ij}(\x)$ and a collisional part $C_{ij}(\x)$, such that
\begin{equation}
 P_{ij}(\x) \ = \  R_{ij}(\x) \ + \ C_{ij}(\x).
\end{equation}

The radiative part can be expanded further in terms of the Einstein coefficients and the average radiation intensity in the line
\begin{equation}
	R_{ij}(\x) \ = \
	\left\{
	\begin{matrix}
		A_{ij} \ + & \! \! \! B_{ij} \ J_{ij}(\x), \ & \text{for} \ i>j \\
		           & \! \! \! B_{ji} \ J_{ij}(\x), \ & \text{for} \ i<j
	\end{matrix}
	\right .
\end{equation}
where $J_{ij}(\x)$ is the local mean intensity in the spectral range of the transition $ij$. It is computed by averaging the specific monochromatic intensity $I_{\nu}(\x, \n)$ over all directions ($\n$) and integrating it over the line profile $\phi_{\nu}^{ij}(\x)$, such that
\begin{equation}
	J_{ij}(\x) \ = \ \oint \frac{\D\Omega}{4\pi} \int_{0}^{\infty} \D\nu \ \phi_{\nu}^{ij}(\x) \ I_{\nu}(\x, \n).
\label{meanIntensity}
\end{equation}

The collisional part of the transition rates is composed of the collisional rates ($K_{ij}^{p}$) for each collision partner ($p$), weighted by their respective abundances
\begin{equation}
    C_{ij}(\x) = \sum_{p\in \mathcal{C}} K_{ij}^{p}(\x) \ n^{p}(\x),
\end{equation}
where $\mathcal{C}$ is the set of collision partners. The position dependence in the collisional rates stems from their dependence on the local temperature of the gas species.

For the models in this paper, we used the energy levels, Einstein coefficients and collisional de-excitation rates from the Leiden Atomic and Molecular Database\footnote{Database can be found at \href{https://home.strw.leidenuniv.nl/~moldata/}{home.strw.leidenuniv.nl/\textasciitilde{}moldata}.} \citep[LAMDA,][]{Schoier2005}. The collisional excitation rates are computed from the de-excitation rates, assuming the detailed balance relation
\begin{equation}
    K_{ji}(\x) \ = \ K_{ij}(\x) \ \frac{g_{j}}{g_{i}} \ \exp \left( \frac{h\nu_{ij}}{\kb T(\x)} \right)
\end{equation}
where $g_{i}$ denotes the statistical weight of the respective level.

We assume the radiative time scales to be much smaller than any other time scale in the system. Hence, to find the level populations given the radiation field, we solve equation (\ref{eq:levelpop}) in the static limit, i.e. assuming that for all levels $i$, $\partial n_{i}(\x) / \partial t = 0$. Dropping the position dependence on all variables, the resulting linear equation for each level $i$ can be written as
\begin{equation}
\begin{split}
	&\sum_{j, \; j<i} \Big\{ \ n_{i} A_{ij} \ - \ \left( n_{j} B_{ji} - n_{i} B_{ij} \right)  J_{ij} \Big\} \\
	&- \sum_{j, \; j>i} \Big\{ \ n_{j} A_{ji} \ - \ \left( n_{i} B_{ij} - n_{j} B_{ji} \right)  J_{ij} \Big\} \\
	&+ \sum_{j=1}^{N} \left\{ \ n_{i} C_{ij} \ - \ n_{j} C_{ji} \right\} \ = \ 0.
\end{split}
\label{eq:stateq}
\end{equation}
It is important to remember that the radiation field, and thus $J_{ij}$, depends on the level populations through the line contributions to the emissivity and opacity (see equation \ref{eq:lines}). This dependence can be expressed mathematically using a Lambda operator. We define this operator such that it yields the mean intensity in the line when acting on the set of all level populations $\textbf{N} \equiv \left\{n_{i}(\x), \ \text{for all } \x \text{ and } i \right\}$.
\begin{equation}
	J_{ij} \ = \  \Lambda_{ij} [\textbf{N}]
\label{eq:Lambda}
\end{equation}
In most practical cases it is unfeasible to directly invert the Lambda operator (i.e. directly solve the radiative transfer equation (\ref{eq:RTE})) and solve the kinetic rate equations (\ref{eq:stateq}) for the level populations. Instead, we solve equation (\ref{eq:stateq}) iteratively, by evaluating $J_{ij}$ using the values from the previous iteration. However, this method, known as Lambda iteration, converges notoriously slowly. Over the years, various methods have been devised to accelerate its convergence  \citep[for an overview see e.g.][and the references there]{Hubeny2014}. We use the operator splitting method \citep{Cannon1973_angle, Cannon1973_frequency} in a very similar way to \cite{Rybicki1991}. The idea is to split the Lambda operator into an approximated part ($\Lambda_{ij}^{\ast}$) that can easily be evaluated and inverted given the current level populations, and a residual part ($\Lambda_{ij}-\Lambda_{ij}^{\ast}$) that can easily be evaluated using the populations of the previous iteration. Hence,
\begin{equation}
    J_{ij} \ = \ \Lambda_{ij}^{\ast} [\textbf{N}] \ + \ \left(\Lambda_{ij} \ - \ \Lambda_{ij}^{\ast}\right) [\textbf{N}^{\dagger}],
\label{eq:Jeff}
\end{equation}
where the dagger ($\dagger$) indicates that the quantity is evaluated using the previous iteration. In this way, the contribution of the level populations of the previous iteration can be minimised. The kinetic rate equations (\ref{eq:stateq}) can thus be rewritten as
\begin{equation}
\begin{split}
	&\sum_{j, \; j<i} \Big\{ \ n_{i} A_{ij} \ - \ \left( n_{j} B_{ji} - n_{i} B_{ij} \right)  \left(\Lambda^{\ast}_{ij}[\textbf{N}] \ + \ J^{\text{eff}}_{ij} \right) \Big\} \\
	&- \sum_{j, \; j>i} \Big\{ \ n_{j} A_{ji} \ - \ \left( n_{i} B_{ij} - n_{j} B_{ji} \right)  \left(\Lambda^{\ast}_{ij}[\textbf{N}] \ + \ J^{\text{eff}}_{ij} \right) \Big\} \\
	&+ \sum_{j=1}^{N} \left\{ \ n_{i} C_{ij} \ - \ n_{j} C_{ji} \right\} \ = \ 0,
\end{split}
\label{eq:ALI}
\end{equation}
where we introduced the effective mean intensity in the line,
\begin{equation}
    J^{\text{eff}}_{ij} \ = \ \left(\Lambda_{ij} \ - \ \Lambda_{ij}^{\ast}\right) [\textbf{N}^{\dagger}].
\end{equation}
Note that the effective mean intensity is now the only quantity that is evaluated using the level populations of the previous iteration. Clearly, the choice of the approximated Lambda operator (ALO) is essential for the success of this acceleration scheme. In some cases the diagonal part of the Lambda operator already suffices \citep{Olson1986}, however in 3D models, a non-local ALO is often preferred.  We discuss our implementation of the ALO, following \cite{Rybicki1991}, in Section \ref{subsec:ALI}.

\section{Numerical implementation}
\label{sec:numerical_implementation}

\subsection{Discretisation of the model}
\label{subsec:discretisation of the model}

The first step in simulating an astrophysical object is finding a way to represent the object on a computer. This comes down to finding an appropriate discretisation scheme for all physical parameters of the model. For radiative transfer simulations the spatial, spectral and directional discretisation schemes are most crucial.

\subsubsection{Spatial discretisation}
\label{subsubsec:spatial}

There are many different types of spatial discretisation schemes, each tailored to their specific use cases. Over the years, there has been a clear evolution from structured schemes, like e.g. regular Cartesian grids, to unstructured schemes, like e.g. Voronoi grids or smoothed particle hydrodynamics (SPH) discretisations.

Since we aim to build a general-purpose library that can easily be integrated with other codes, we do not want to tie \Magritte{} to a certain discretisation scheme. Instead, we designed our algorithms such that they only require data that can easily be deduced from any discretisation scheme, and yet allow us to efficiently trace rays and solve the transfer equation. \Magritte{}'s ray tracing algorithm, presented in Section \ref{subsec:ray_tracing}, only requires the positions of the cell centres\footnote{We do not require a strict definition of the cell centre. If we define a cell as a unit in the discretisation of the spatial volume, then the cell centre may be any point in that volume. (We only use the cell centre to locate the cell.)} (or equivalently the positions of the particles in an SPH scheme) and the nearest neighbours lists for each cell (or particle). Hence, the input is effectively a point cloud complemented with nearest neighbour information.
The boundary of the model can then be defined as the set of points in the convex hull of the point cloud.

In principle, one could use a separate spatial discretisation to sample the density, velocity and temperature distributions of the model. In practice, however, one usually samples all three on the same discretisation. This is the case, especially for hydrodynamics computations, where all these parameters should be sampled equally well. However, for radiative transfer computations, especially when considering lines, it is essential to properly sample any changes in the velocity field along a certain line of sight. Since this effect depends on the velocity field in a certain direction it is difficult to fully take this into account in the spatial discretisation. In \Magritte{}, when detecting large changes in the velocity field along a ray, we make an appropriate interpolation on-the-fly when ray-tracing, without adjusting the mesh (see also Section \ref{subsec:ray_tracing}).

\subsubsection{Spectral discretisation}
\label{subsubsec:spectral}

The requirements for the spectral discretisation vary for different stages in the computation. For instance, when determining the level populations, we are only interested in the radiation in the lines, whereas when computing spectra we require a proper frequency sampling over the full spectrum. To accommodate this, \Magritte{} can change its spectral discretisation throughout a simulation.

At the stage where the level populations are obtained, the frequency bins are distributed to suit the integration of the radiation field over the line. In \Magritte{} these integrals are evaluated using quadrature formulae. Assuming a Gaussian line profile, the corresponding Gauss-Hermite quadrature for any frequency dependent function $y(\nu)$ is given by
\begin{equation}
\int_{0}^{\infty} \D\nu \ \phi_{\nu}^{ij}(\x) \ y_{\nu} \ \rightarrow \ \sum_{n=1}^{N_{\text{q}}} \ w_{n} \ y \left(\nu_{ij} + r_{n} \delta\nu_{ij}(\x)\right),
\end{equation}
where $N_{\text{q}}$ is the number of quadrature points and the quadrature weights are given by
\begin{equation}
w_{n} \ = \ \frac{2^{N_{\text{q}}-1}N_{\text{q}}!}{\big( N_{\text{q}} H_{N_{\text{q}}-1}(x_{n}) \big)^{2}},
\end{equation}
$H_{N_{\text{q}}-1}$ is the physicists' version of the Hermite polynomial and the $r_{n}$ are the roots of the physicists' version of the Hermite polynomial $H_{N_{\text{q}}}(x)$ \citep[see e.g.][]{Abramowitz1972}. To be able to easily evaluate these quadratures in \Magritte{}, we define a separate set of frequency bins for each cell, given by
\begin{equation}
  \left\{\nu_{ij} + r_{n} \, \delta\nu_{ij}(\x), \text{ for each transition $ij$ and root $r_{n}$} \right\},
\end{equation}
possibly appended with additional frequency bins. Note that this set has to be different for each cell, since it depends on the local line profile width $\delta\nu_{ij}(\x)$.

At the stage where spectra or images are created, extra frequency bins can be appended to the list above to improve the sampling of the spectrum. The current version of \Magritte{} allows one to append a set of user-defined frequency bins, or to add extra bins with a user-defined spacing around each line.

\subsubsection{Directional discretisation}
\label{subsubsec:directional}

\Magritte{} is a ray-tracing code, i.e. the radiation field is determined by solving the radiative transfer equation along a set of rays (straight lines) originating from each cell centre. A ray can be defined by a point, in our case the cell centre, and a direction. The direction of the rays will play a key role in scattering and will determine the viewing angles for the images we can take.

In general, there are no preferred directions. Therefore, we discretise the directions uniformly. In 1D and 2D models this is trivial. In 3D, we determine the direction of a ray using the \textsc{HEALPix}\footnote{\label{HEALPix}Source code for \textsc{HEALPix} can be found at \href{http://healpix.sourceforge.net}{healpix.sourceforge.net}.} discretisation of the sphere \citep{Gorski2005}. Given a level of refinement, $\ell$, it discretises a unit sphere in $N_{\text{rays}}=12 \times 4^{\ell}$ uniformly distributed pixels of equal area. For each pixel, there is an associated unit vector pointing from the origin of the sphere to the pixel centre. These unit vectors determine the directions of \Magritte{}'s rays for 3D simulations. Hence, a directional average for a quantity $y(\n)$ can be translated into an average over rays,
\begin{equation}
    \frac{1}{4\pi} \oint \D \Omega \ y(\n) \ \rightarrow \ \frac{1}{N_{\text{rays}}} \sum_{r=1}^{N_{\text{rays}}} \ y_{r} .
\end{equation}

The uniform directional sampling scheme bears the danger of missing the contributions of very localised sources of emissivity or opacity. Furthermore, there might be situations in which there is one or more preferred directions, and one might better consider a non-uniform distribution of the ray directions. Therefore, in future versions, we will investigate more advanced directional weighting schemes. In any case, the internal structure of \Magritte{} allows for any distribution of rays, allowing us to easily explore various directional distribution schemes in the future.

\subsection{Ray tracing}
\label{subsec:ray_tracing}

In order to solve the radiative transfer equation along a certain ray, the emissivity and opacity of the cells that are encountered along that ray must be known. Furthermore, the path length that the ray traces through each cell must be computed. All this must be done assuming only a point cloud with nearest neighbour information.

\begin{figure}
	\centering
	\includegraphics[width=\columnwidth]{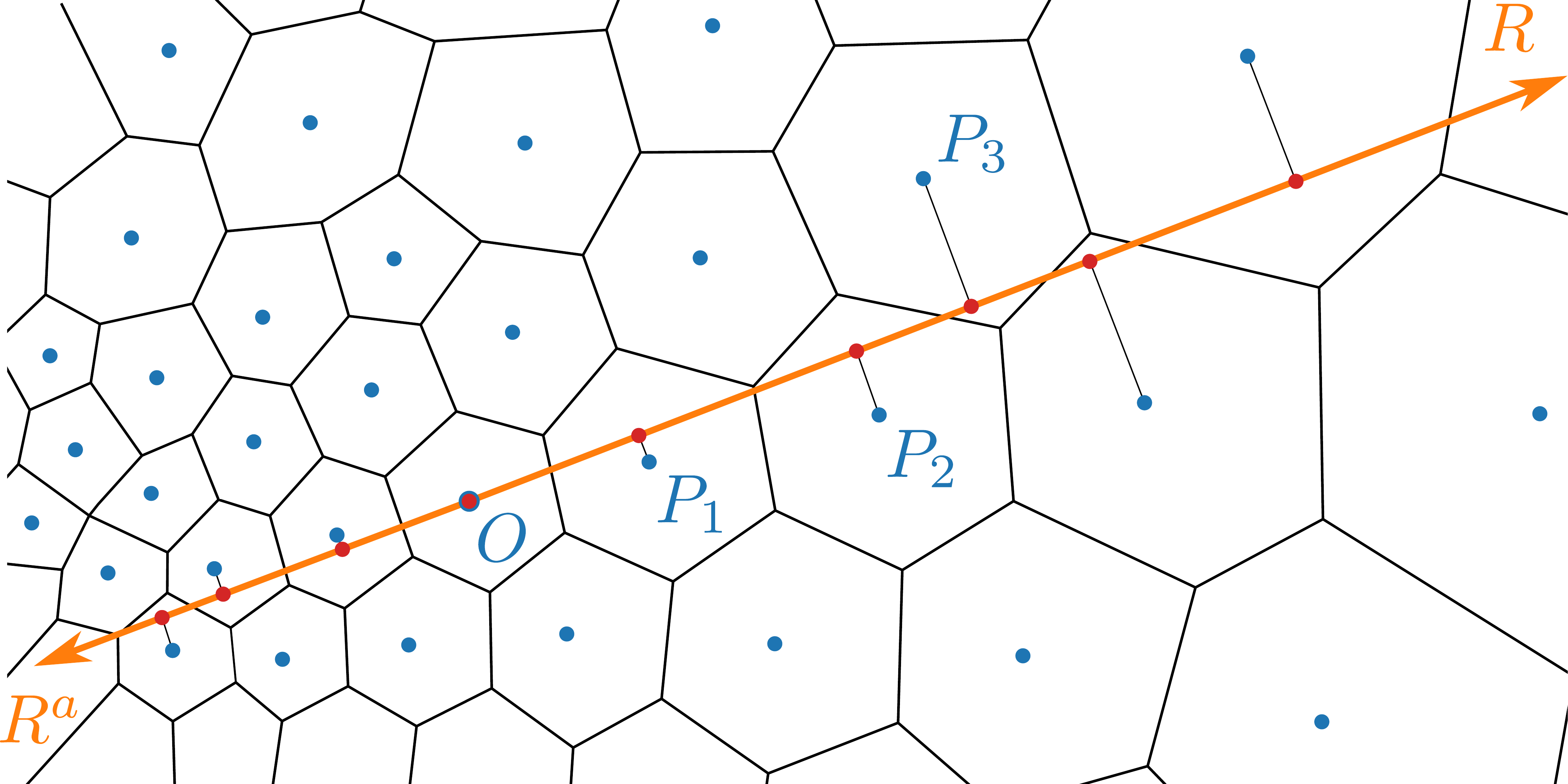}
  \caption{A visual representation \Magritte{}'s ray tracing algorithm for the ray $R$ originating from cell $O$. The goal is to find which cells are encountered along the ray and hence which cell centres should be projected on the ray. Clearly, the cell $O$ itself lies on the ray. The next cell  encountered is the neighbour of $O$ that lies closest to the ray. We call this cell $P_{1}$. Now the next cell to be projected is the neighbour of $P_{1}$ that lies closest to the ray and that is further away from $O$ than $P_{1}$. The last condition is there to ensure that one proceeds along the ray towards the boundary. This process is repeated until the boundary of the mesh is reached.}
  \label{fig:ray_tracing}
\end{figure}

The idea of \Magritte{}'s ray tracing algorithm is to walk along the ray from one cell to the next and determine the path length through each cell by projecting the cell centres onto the ray. To determine which cell is next, the set of all nearest neighbours of the current cell is considered. From this set the neighbour is chosen which is closest to the ray and whose projection on the ray lies farther than that of the current cell. This procedure is then repeated until the boundary of the mesh is reached. Figure \ref{fig:ray_tracing} shows a visual example of how this algorithm works. Once the rays have been traced, the transfer equation can be solved.

In each step from one cell to the next, the change in velocity along the ray is computed and checked for large variations. If the velocity field, and thus the resulting Doppler shift, changes too much the emissivity and opacity are interpolated between the cells such that the velocity steps are only a certain (user defined) fraction of the local line width. In this way, we avoid losing or improperly accounting for line contributions due to an inadequate sampling of the velocity field.

\subsection{Solving the transfer equation along a ray pair}
\label{subsec:solving_the_transfer_equation}

In \Magritte{}, the radiation field is obtained by solving the radiative transfer equation along each pair of a ray and its antipode through the model. Although in this paper we are only concerned with line radiative transfer, we present our general solution method for the full radiative transfer equation (\ref{eq:RTE}) including scattering. In this way, the treatment in this paper more closely resembles the actual implementation in the code and paves the way for our future work.

For numerical stability, we solve the transfer equation in its second-order form as suggested by \cite{Feautrier1964}. We define the mean intensity-like ($u$) and flux-like quantity ($v$) along a ray as
\begin{equation}
\begin{split}
		u_{\nu}(\n) \ &\equiv \ \frac{1}{2}\Big( I_{\nu}(\n) \ + \ I_{\nu}(-\n) \Big) , \\
		v_{\nu}(\n) \ &\equiv \ \frac{1}{2}\Big( I_{\nu}(\n) \ - \ I_{\nu}(-\n) \Big) ,
\end{split}
\end{equation}
to describe the radiation field. To simplify notation further on, we also define new quantities to represent the scattering redistribution function and scattering opacity up and down the ray
\begin{equation}
\begin{split}
	\Phi^{\pm}_{\nu \, \nu'}(\n \, ,\n') \ &\equiv \ \frac{1}{2} \Big( \Phi_{\nu \, \nu'}(\n \, ,\n') \ \pm \ \Phi_{\nu \, \nu'}(-\n \, ,\n') \Big) , \\
	\chi^{\pm}_{\nu}(\n) \ &\equiv \ \frac{1}{2} \Big( \chi^{\text{sca}}_{\nu}(\n) \ \pm \ \chi^{\text{sca}}_{\nu}(-\n) \Big) .
\end{split}
\end{equation}
Finally, to avoid lengthy integral equations we define
\begin{equation}
	\Psi^{\pm}_{\nu}\left(\n\right) \ \equiv \ \oint \D\Omega' \int_{0}^{\infty} \D \nu' \ \Phi^{\pm}_{\nu \, \nu'}(\n \, ,\n') \ I_{\nu'}(\n') .
\end{equation}
From here onward, we drop all $\nu$ and $\n$ dependencies for notational simplicity. We proceed by adding and subtracting the transfer equation (\ref{eq:RTE}), once for $\n$ and once for $-\n$. This yields a coupled set of first-order differential equations in $u$ and $v$,
\begin{equation}
\begin{split}
	\n \cdot \nabla v \ &= \ - \left(\chi + \chi^{+}\right) u \ - \ \chi^{\text{-}} \ v \ + \ \Psi^{+} \ + \ \eta , \\
	\n \cdot \nabla u \ &= \ - \left(\chi + \chi^{+}\right) v \ - \ \chi^{\text{-}} \ u \ + \ \Psi^{-} . \\
\end{split}
\label{eq:RTE2}
\end{equation}
Solving the equations, once for $u$ and once for $v$, yields a set of second-order differential equations, which are our generalised versions of the Feautrier equations \citep{Feautrier1964}
\begin{equation}
\begin{split}
	\left(1-\mathcal{D}^{2}\right) \ u \ &= \ \frac{\eta + \Psi^{+}}{\chi + \chi^{+}} \ - \ \mathcal{D}\left(\frac{\Psi^{-}}{\chi + \chi^{+}}\right),  \\
	\left(1-\mathcal{D}^{2}\right) \ v \ &= \ \frac{\Psi^{-}}{\chi + \chi^{+}} \ - \ \mathcal{D}\left(\frac{\eta + \Psi^{+}}{\chi + \chi^{+}}\right),
\end{split}
\label{eq:GFE}
\end{equation}
where we defined a new differential operator $\mathcal{D}$ as
\begin{equation}
\mathcal{D} \ \equiv \ \frac{1}{\chi + \chi^{+}} \left(\chi^{-} \ + \ \n \cdot \nabla \right).
\end{equation}
Note that the order of the factors in the definition of $\mathcal{D}$ is important, since they do not commute. Both equations in (\ref{eq:GFE}) are still coupled through their $\Psi^{\pm}$ terms. However, the contributions between the scattering opacity, $\chi^{\text{sca}}_{\nu}(\n)$, and the scattering redistribution function, $\Phi_{\nu \, \nu'}(\n \, ,\n')$, can be arranged such that $\Phi_{\nu \, \nu}(\n \, , \n) = 0$. Hence, the coupling between the equations in (\ref{eq:GFE}) can be weakened.

The generalised Feautrier equations (\ref{eq:GFE}) can thus be solved in an iterative way by evaluating their right hand sides using the solution of the previous iteration. In each iteration, two separate ordinary second-order differential equations then have to be solved. The boundary conditions can be determined from the incoming radiation on both sides of the ray pair. By making a Taylor expansion of the intensity, the incoming radiation can be related to $u$ and $v$, in the same way as in the standard Feautrier procedure with improved boundary conditions for plane parallel geometries by \cite{Auer1967}.

In \Magritte{}, the radiation field  is computed by solving the equations (\ref{eq:GFE}) for each ray pair. Since the right hand sides are treated in an iterative way, these act effectively as sources. The second-order differential operators on the left hand sides will result in tridiagonal matrices on the discretised ray pairs. The form of the equations in (\ref{eq:GFE}) still resembles the original Feautrier equations enough that the standard solution method \citep{Feautrier1964} with the numerical improvements by \cite{Rybicki1991} can readily be adapted to this generalised case.

\subsection{Accelerated Lambda Iteration}
\label{subsec:ALI}

From the first equation in (\ref{eq:GFE}) we can identify the Lambda operator for our solution scheme, as defined in equation (\ref{eq:Lambda})
\begin{equation}
  \Lambda_{ij} [\textbf{N}] \ = \ \mathcal{L}_{ij} \left[\frac{\eta + \Psi^{+}}{\chi + \chi^{+}} \ - \ \mathcal{D}\left(\frac{\Psi^{-}}{\chi + \chi^{+}}\right) \right],
\label{eq:ourLambda}
\end{equation}
where we used the auxiliary linear operator $\mathcal{L}_{ij}$, defined as
\begin{equation}
  \mathcal{L}_{ij} [\, . \, ] \ = \ \frac{1}{2} \ \oint \frac{\D\Omega}{4\pi}  \int_{0}^{\infty} \D \nu \ \phi_{\nu}^{ij} \ \left(1-\mathcal{D}_{\nu}^{2}\right)^{-1} [\, . \,].
\label{eq:L_operator}
\end{equation}
Following \cite{Rybicki1991}, we can construct an approximation to the Lambda operator by considering only the diagonal band of the matrix representation of the auxiliary operator $\mathcal{L}_{ij}$. We call this operator $\mathcal{L}_{ij}^{\ast}$. The operator $\mathcal{L}_{ij}^{\ast}$ is easy enough to invert, due to its band diagonal structure. However, using it as the ALO would render (\ref{eq:ALI}) into a system of non-linear equations for the level populations, which would still be hard to solve. In order to retain the linearity of (\ref{eq:ALI}), we instead define our ALO as
\begin{equation}
  \Lambda_{ij}^{\ast} [\textbf{N}] \ = \  \frac{n^{\dagger}_{j} B_{ji} - n^{\dagger}_{i} B_{ij}}{n_{j} B_{ji} - n_{i} B_{ij}} \ \mathcal{L}_{ij}^{\ast} \left[\frac{\eta^{ij}(\textbf{N})}{\chi(\textbf{N}^{\dagger}) + \chi^{+}(\textbf{N}^{\dagger})} \right]
\label{eq:ALO}
\end{equation}
where we only evaluate the line emissivity in the argument of $\mathcal{L}_{ij}^{\ast}$ with the new level populations and add an extra factor which goes to unity when the level populations converge. Since the line emissivity is linear in the level populations (see equation \ref{eq:lines}) the statistical equilibrium equation will remain linear in the new level populations and can be written as
\begin{equation}
\begin{split}
	&\sum_{j, \; j<i} \Big\{ \ n_{i} A_{ij} \ - \ \tilde{\Lambda}_{ij}^{\ast}[n_{i}] \ - \ \left( n_{j} B_{ji} - n_{i} B_{ij} \right)  J^{\,\text{eff}}_{ij} \ \Big\} \\
	&-\sum_{j, \; j>i} \Big\{ \ n_{j} A_{ji} \ - \ \tilde{\Lambda}_{ji}^{\ast}[n_{j}] \ - \ \left( n_{i} B_{ij} - n_{j} B_{ji} \right)  J^{\,\text{eff}}_{ij} \ \Big\} \\
	&+ \sum_{j=1}^{N} \left\{ \ n_{i} C_{ij} \ - \ n_{j} C_{ji} \right\} \ = \ 0,
\end{split}
\label{eq:ourALI}
\end{equation}
where the effective mean intensity, defined in (\ref{eq:Jeff}), is given by
\begin{equation}
  J_{ij}^{\text{eff}} \ = \  J_{ij} \ - \ \mathcal{L}_{ij}^{\ast} \left[\frac{\eta^{ij}(\textbf{N}^{\dagger})}{\chi(\textbf{N}^{\dagger}) + \chi^{+}(\textbf{N}^{\dagger})} \right]
\label{eq:ourJeff}
\end{equation}
and we introduced another auxiliary approximated operator
\begin{equation}
  \tilde{\Lambda}_{ij}^{\ast} [n_{i}] \ = \ \frac{h}{4\pi}  \left(n^{\dagger}_{j} B_{ji} - n^{\dagger}_{i} B_{ij}\right) \ \mathcal{L}_{ij}^{\ast} \left[\frac{A_{ij} \ \nu \phi^{ij}_{\nu} \ n_{i}}{\chi(\textbf{N}^{\dagger}) + \chi^{+}(\textbf{N}^{\dagger})} \right].
\end{equation}
Note that this operator is linear in the new level populations and not symmetric in the level indices $i$ and $j$. However, since our ALO (\ref{eq:ALO}) is symmetric in these indices, it can be implemented on the level of the transition matrix.

Apart from the ALO, we also use Ng-acceleration \citep{Ng1974} in \Magritte{} to speed-up convergence even more. This acceleration method introduces a special iteration step every $M$ number of (regular)
iterations. In this special iteration step, the level populations of the next iteration are predicted by a linear combination of
the populations of the previous $M$ iterations. This is done by minimising the change in the level populations for the prediction
based on the last $M-1$ and the one to last $M-1$ iterations. Since the Ng-iteration step does not
require the computation of the radiation field, it is much faster than a regular iteration and thus accelerates the iteration process. The Ng-method allows us to specify a weight for the contribution of the different levels to the prediction \citep[see e.g.][]{Olson1986}. For this paper, a uniform weighting scheme was applied, but \Magritte{} can readily be adapted to handle any other scheme.

\subsection{Imaging the model}
\label{subsec:imaging_the_model}

When modelling astrophysical objects in 3D, one often requires images of the model from several viewpoints in several frequency ranges, in order to compare the model with observations.

In \Magritte, these images can be obtained using the solution of the outward directed radiation field on the endpoints of each ray pair. One can construct the image by considering the outward directed radiation along a certain ray and projecting the locations of the originating points on the plane orthogonal to the ray. The result is a set of points on a plane with a corresponding intensity, which can be easily rendered into an image. Note that since every point in the model contributes to one point in the image, the spatial resolution of the resulting image is exactly equal to the highest achievable resolution for that model.

\section{Tests \& benchmarks}
\label{sec:tests_and_benchmarks}

To demonstrate the validity of our methods and to better understand their limitations, we have conducted a series of comparisons with analytical models and benchmarked against established radiative transfer codes. The analysis for these tests and benchmarks was performed in a collection of \textsc{Jupyter} notebooks \citep{Kluyver2016}, which are publicly available on GitHub\footnote{\label{benchmarks}Benchmarks can be found at \href{https://github.com/Magritte-code/Benchmarks}{github.com/Magritte-code/Benchmarks}.}.

\subsection{Semi-analytical tests}

To assess the accuracy of \Magritte{}'s ray tracer and radiative transfer solver, we first reproduce some semi-analytically solvable line radiative transfer models. This will help us later to better assess the uncertainties associated to the results of our simulations. We cannot overemphasise the importance of these analytical tests as they are the only way to obtain absolute measures of the accuracy.

\subsubsection{Homogeneous Hubble-Lema\^itre models}

As a first test, we consider the radiative transfer of a single line on a uniformly spaced grid with a constant molecular abundance and temperature distribution, and a constant velocity gradient. The velocity distribution is thus given by the Hubble-Lema\^itre law
\begin{equation}
    \varv(r) \ = \ c \Delta \beta \ r ,
\end{equation}
where we parametrised the velocity gradient $\Delta \beta$ as a fraction of the speed of light $c$.
The boundary condition is given by incoming cosmic microwave background (CMB) radiation, i.e a black-body spectrum $B_{\nu}$ of temperature $T_{\text{CMB}} = 2.725$ K. If we assume LTE level populations, the line source function $S_{\nu_{ij}}$ is spatially constant. In that case, one can find the mean intensity by directly integrating the transfer equation, yielding
\begin{equation}
  J_{\nu}(\x) \ = \ S_{\nu_{ij}} \ + \ \left(B_{\nu}-S_{\nu_{ij}}\right) \ \oint \frac{\D \Omega}{4\pi} \ e^{-\tau_{\nu}\left(\x, \n\right)}
\label{eq:J_for_constant_S}
\end{equation}
where the optical depth, assuming Gaussian line profiles centred around $\nu_{ij}$ and with a line profile width $\delta\nu_{ij}$, is given by
\begin{equation}
  \tau_{\nu}(\ell) \ = \ \frac{\chi_{ij}}{2 \nu \Delta \beta} \left\{ \text{Erf}\left[\frac{\nu-\nu_{ij}}{\delta\nu_{ij}}\right] \ - \ \text{Erf}\left[\frac{\nu\left(1-\Delta \beta \ell\right) -\nu_{ij}}{\delta\nu_{ij}} \right] \right\},
\label{eq:optical_depth}
\end{equation}
where $\text{Erf}$ is the error function, and $\ell(\x, \n)$ is the distance from point $\x$ to the boundary, as measured along the ray in direction $\n$. Since the Hubble-Lem\^itre velocity law is both translation and rotation invariant, only the total distance to the boundary appears in the expression for the optical depth.

Considering only a single ray in the interval  $[-R,R]$, the mean intensity in $r\in[0,R]$, as expressed in equation (\ref{eq:J_for_constant_S}), reads
\begin{equation}
    J_{\nu}(r) \ = \ S_{\nu_{ij}} \ + \ \frac{1}{2} \left(B_{\nu}-S_{\nu_{ij}}\right) \ \left[e^{-\tau_{\nu}(r)} + e^{-\tau_{\nu}(R-r)}\right],
\label{eq:Hubble-Lemaitre1D}
\end{equation}
where the average over all directions reduces to the average intensity flowing up and down the ray.

In three dimensions, assuming a spherical boundary with radius $R$, the mean intensity expressed in equation (\ref{eq:J_for_constant_S}) reduces to
\begin{equation}
J_{\nu}(r) \ = \ S_{\nu_{ij}} \ + \ \frac{1}{2} \left(B_{\nu}-S_{\nu_{ij}}\right) \ \int_{0}^{\pi} \D \theta \ \sin \theta \ e^{-\tau_{\nu}\left(\ell(r, \theta)\right)}
\label{eq:Hubble-Lemaitre3D}
\end{equation}
where the distance to the boundary $\ell(r, \theta)$ is given by
\begin{equation}
\ell(r,\theta) \ = \ r \cos \theta \ + \ \sqrt{R^{2} - r^{2} \sin^{2} \theta} .
\end{equation}
The $\theta$-integral in the expression for the mean intensity can easily be computed numerically. Note that introducing the spherical boundary breaks the translation invariance of the problem.

Although these are simple models, they can demonstrate some key issues in numerical radiative transfer modelling. In particular, both models can be used to directly assess the accuracy of the radiative transfer solver and to test the sampling in velocity space. Especially in line radiative transfer it is crucial to properly sample the velocity field, since too large a step in velocity from one cell to the next can Doppler-shift a line directly from one wing to the other without capturing the effect of the core of the line. This can be tested by adjusting the velocity gradient. By considering both the single ray and full 3D model we can also assess the quality of the spatial interpolations onto the rays.

For this test we used a fictitious 2-level species in a (radially) uniformly spaced grid $[-R,R]$ with $R=495$ km, and with a velocity gradient $c \Delta\beta = 0.01$ s$^{-1}$. The line data for the fictitious 2-level species are summarised in Table \ref{tab:test_species}. We assume a constant H$_{2}$ number density of $n^{\text{H}_{2}} = 1.0 \times 10^{12}$ m$^{-3}$ and a constant fractional abundance of the fictitious 2-level species $X \equiv n^{\text{fict}}/n^{\text{H}_{2}} = 10^{-4}$. To obtain the level populations, we assume LTE with a constant temperature distribution $T = 45$ K. Furthermore, we assume  the gas has no turbulent velocity component. The 3D model is obtained from the 1D model by mapping each 1D grid point to a shell of 3D grid points uniformly distributed over a sphere. The model parameters for \Magritte{} can be found in Table \ref{tab:test_parameters}.

\begin{figure}
	\centering
	\includegraphics[width=\columnwidth]{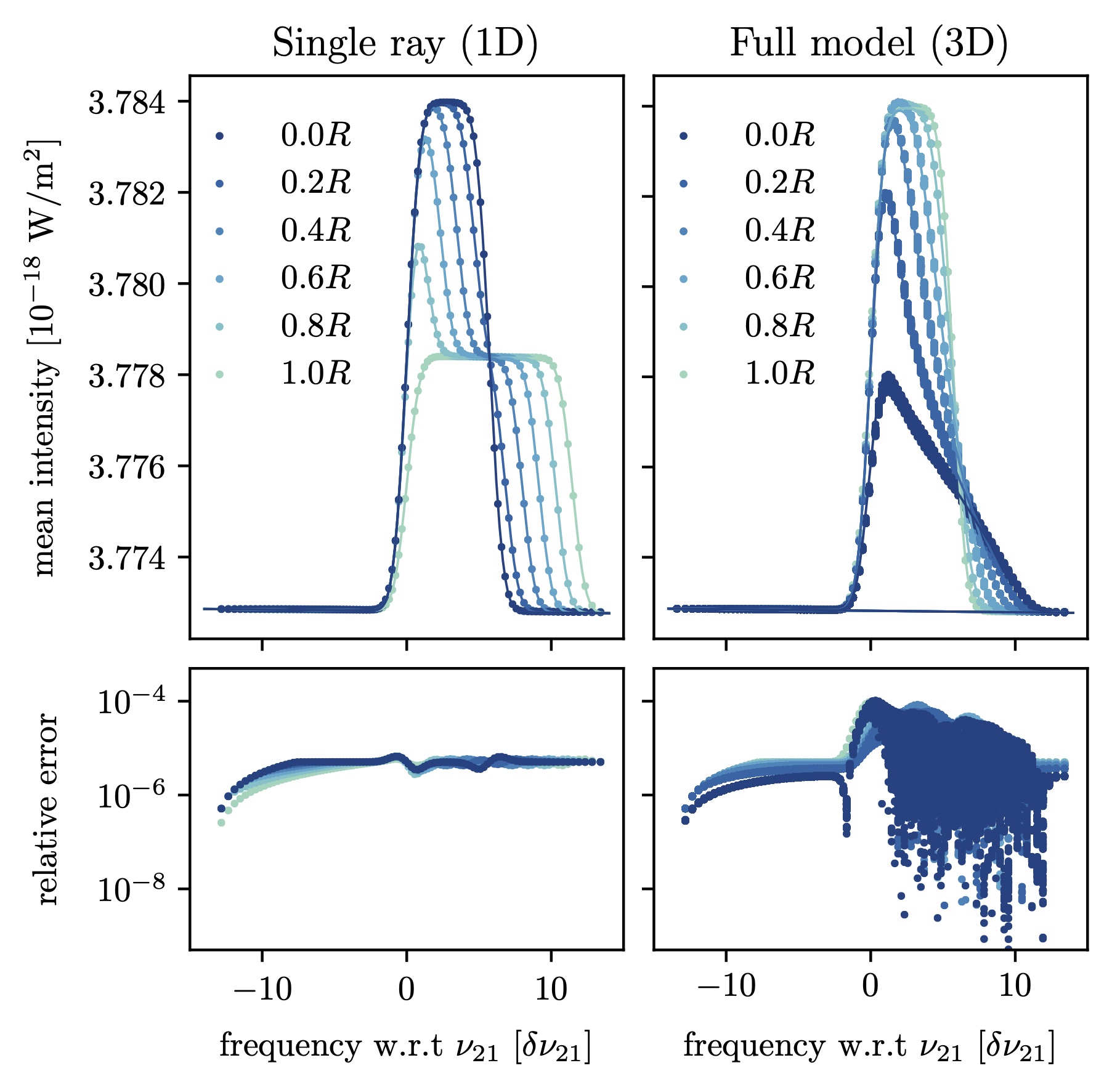}
  \caption{Comparison between \Magritte{} and the semi-analytical solution of the mean intensity as a function of frequency in the Hubble-Lema\^itre model, evaluated at different radii. The dots indicate \Magritte{}'s results and the line represent the analytic results. Frequencies are expressed with respect to the line centre $\nu_{21} \approx 179.88$ GHz as a fraction of the line profile width $\delta\nu_{21} \approx 519.03$ kHz. The relative error of two values is measured as twice the absolute difference over their sum.}
  \label{fig:Hubble-Lemaitre}
\end{figure}

Figure \ref{fig:Hubble-Lemaitre} shows a comparison between the solution of \Magritte{} and the semi-analytical solutions (\ref{eq:Hubble-Lemaitre1D}, \ref{eq:Hubble-Lemaitre3D}) of the Hubble-Lema\^itre models. \Magritte{}'s numerical result clearly agrees with the analytic solution with a relative error well below $10^{-4}$ almost everywhere, where the relative error of two values is measured as twice the absolute difference over their sum.

\begin{table}
	\centering
	\caption{Line data of the fictitious 2-level species. This is the same fictitious 2-level species as used in Problem 1 in \protect\cite{vanZadelhoff2002}.}
	\label{tab:test_species}
	\begin{tabular}{c c c c}
	    \hline
        $E_{2}-E_{1}$ [cm$^{-1}$] & $g_{2}/g_{1}$ &   $A_{21}$ [ s$^{-1}$] & $K_{21}^{\text{H}_{2}}$ [cm$^{3}$s$^{-1}$] \\ \hline
        6.0                     & 3.0             & $1.0 \times 10^{-4}$   & $2.0 \times 10^{-10}$
	\end{tabular}
\end{table}

\subsubsection{Simple power-law density distribution}
\label{subsubsec:density_distribution}

As a second semi-analytic test, we consider the radiative transfer of a single line on a logarithmically spaced grid, with a constant temperature distribution, with no velocity field, and a (spherically) symmetric density distribution given by a power-law
\begin{equation}
    n^{\text{H}_{2}}(r)  \ = \
\begin{cases}
    \ 0 & \text{for} \ r < R_{\text{in}} \\
    \ n^{\text{H}_{2}}\left(R_{\text{in}}\right) \ \left(\frac{R_{\text{in}}}{r}\right)^{2} & \text{for} \ r \geq R_{\text{in}}
\end{cases}
\label{eq:density_distribution}
\end{equation}
where $r_{\text{in}}$ is the inner radius of the model. The boundary condition is again given by incoming cosmic microwave background (CMB) radiation, i.e a black-body spectrum $B_{\nu}$ of temperature $T_{\text{CMB}} = 2.725$ K. If we again assume LTE level populations, the line source function $S_{\nu_{ij}}$ is spatially constant. As a result the mean intensity is again given by equation (\ref{eq:J_for_constant_S}). To compute the optical depth, one needs to integrate the density distribution along every ray. Assuming a spherical boundary with radius $R$, the optical depth is given by
\begin{equation}
    \tau_{\nu}(r,\theta) \ = \ \frac{\chi_{ij} \ \phi^{ij}_{\nu} \ r}{\sin \theta} \left(\frac{\pi}{2} - \theta \ + \ \arccos \left(\frac{r \sin \theta}{R}\right) \ - \ f(r, \theta) \right)
\end{equation}
where the function $f(r,\theta)$ accounts for the rays that go through the empty core ($r<R_{\text{in}}$) of the model and is given by
\begin{equation}
\begin{split}
    f\left(r, \theta\right) \ = \
\begin{cases}
    \ 2 \arccos \left(\frac{r \sin \theta}{R_{\text{in}}}\right)  & \ \text{for} \ \ \theta < \theta_{\text{core}} \\
    \ 0 & \  \text{for} \ \ \theta \geq \theta_{\text{core}} \end{cases}
\end{split}
\end{equation}
and whether or not a ray passes through the empty core is determined by the direction of the ray at each radius, $\theta_{\text{core}} = \arcsin \left(R_{\text{in}} / r\right)$.

Considering only a single ray in the interval $[-R,R]$, the mean intensity is given by
\begin{equation}
    J_{\nu}(r) \ = \ S_{\nu_{ij}} \ + \ \frac{1}{2} \left(B_{\nu}-S_{\nu_{ij}}\right) \ \left[e^{-\tau_{\nu}(r,0)} + e^{-\tau_{\nu}(r,\pi)}\right],
\label{eq:density_distribution1D}
\end{equation}
where the average over all directions reduces to the average intensity flowing up and down the ray. Note that one should be careful in taking the limits $\theta \rightarrow 0$ and $\theta \rightarrow \pi$, but that both are well-defined.

In three dimensions, one can simply integrate over the entire solid angle to obtain the mean intensity
\begin{equation}
J_{\nu}(r) \ = \ S_{\nu_{ij}} \ + \ \frac{1}{2} \left(B_{\nu}-S_{\nu_{ij}}\right) \ \int_{0}^{\pi} \D \theta \ \sin \theta \ e^{-\tau_{\nu}\left(r, \theta\right)}.
\label{eq:density_distribution3D}
\end{equation}
However, one should be careful of distinguishing between rays that do and do not pass through the empty core of the model.

For this test we used the same fictitious 2-level species as before (Table \ref{tab:test_species}) in a radially logarithmically spaced grid $[-R,R]$ with $R_{\text{in}} = 1.0 \times 10^{13}$ m and $R=7.8 \times 10^{16}$ m, and without a velocity field. The H$_{2}$ number density just outside the empty core is $n^{\text{H}_{2}}\left(R_{\text{in}}\right) = 2.0 \times 10^{13}$ m$^{-3}$ and a constant fractional abundance of the fictitious 2-level species $X \equiv n^{\text{fict}}/n^{\text{H}_{2}} = 10^{-6}$ is used. To obtain the level populations, we assume LTE with a constant temperature distribution $T = 20$ K. Furthermore, the gas has a turbulent velocity component of $v_{\text{turb}} = 150$ m s$^{-1}$. The 3D model is obtained from the 1D model by mapping each 1D grid point to a shell of 3D grid points uniformly distributed over a sphere. The model parameters for \Magritte{} can be found in Table \ref{tab:test_parameters}. This model setup is identical to Problem 1b in \cite{vanZadelhoff2002}. However, here we are only interested in the resulting radiation field when the levels are in LTE (see also Section \ref{subsec:cross-code_benechmarks}).

\begin{figure}
	\centering
	\includegraphics[width=\columnwidth]{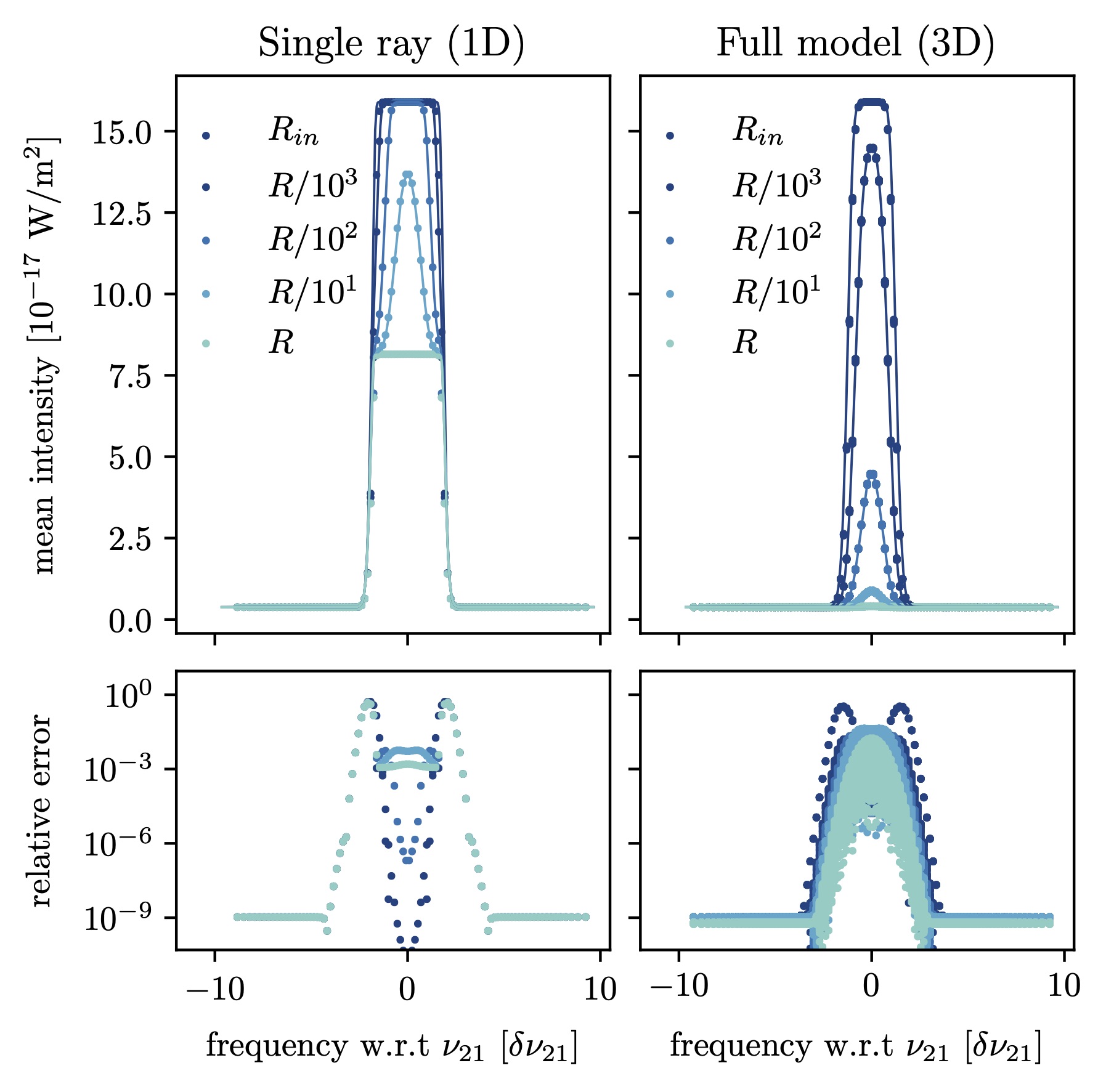}
  \caption{Comparison between \Magritte{} and the semi-analytical solution of the mean intensity as a function of frequency in a model with a simple power-law density distribution, evaluated at different radii. The dots indicate \Magritte{}'s results and the line represent the analytic results. Frequencies are expressed with respect to the line centre $\nu_{21} \approx 179.88$ GHz as a fraction of the line profile width $\delta\nu_{21} \approx 357.53$ kHz. The relative error of two values is measured as twice the absolute difference over their sum.}
  \label{fig:density_distribution}
\end{figure}

Figure \ref{fig:density_distribution} shows a comparison between the solution of \Magritte{} and the semi-analytical solutions (\ref{eq:density_distribution1D}, \ref{eq:density_distribution3D}) of the simple power-law density distribution models. \Magritte{}'s numerical result clearly agrees with the analytic solution. Only at the steep edges of the line is there a larger relative error ($\sim 0.4$), which can be attributed to the steepness of the solution and the discrete mesh.

\begin{table}
\centering
\caption{\Magritte{} parameters for the semi-analytic test models.}
\label{tab:test_parameters}
\begin{tabular}{l | r | r | r | r | r}
  \hline
  model                                 &    & ($N_{\text{shells}}$) & $N_{\text{cells}}$ & $N_{\text{rays}}$ & $N_{\text{q}}$  \\ \hline
  \multirow{2}{*}{Hubble-Lema\^itre}    & 1D & 50 &    100 &   2 & 100 \\
                                        & 3D & 50 & 12 528 & 192 & 100 \\ \hline
  \multirow{2}{*}{density distribution} & 1D & 50 &    100 &   2 & 100 \\
                                        & 3D & 50 & 12 528 & 192 & 100 \\
\end{tabular}
\end{table}

\subsection{Cross-code benchmarks}
\label{subsec:cross-code_benechmarks}

There are no analytic solutions for the full non-LTE line radiative transfer problem, so the only way to fully test \Magritte{}'s line radiative transfer module is by benchmarking it against established codes. Although this does not prove the validity of the code, it is reassuring to find the same results in different ways.

For the benchmarks we used the (1D) problems presented in \cite{vanZadelhoff2002} (from here on referred to as the benchmark paper) and compared our results with the publicly available version of the 1D Monte Carlo radiative transfer code \textsc{Ratran}\footnote{Source code can be found at \href{https://personal.sron.nl/~vdtak/ratran/frames.html}{personal.sron.nl/$\sim$vdtak/ratran/frames.html}.\label{ratran}} \citep{Hogerheijde2000}. Since \Magritte{} is intrinsically multidimensional, the 1D benchmarking models were mapped to their 3D equivalents by mapping each 1D grid point to a shell of 3D grid points uniformly distributed over a sphere.

\subsubsection{Van Zadelhoff Problem 1 a/b}
\label{subsubsec:VZ_p1ab}

The first test, referred to as problem 1 a/b in the benchmark paper, considers a fictitious two-level species in a spherically symmetric cloud, without velocity field, with a constant temperature distribution, and a density distribution given by a power law. The entire model can thus be defined analytically. The model setup is essentially the same as in the simple power-law density distribution test in Section \ref{subsubsec:density_distribution}. The only difference is that in Problem 1a the relative molecular abundance $X=10^{-8}$ results in a low optical depth, whereas in Problem 1b the relative molecular abundance is $X=10^{-6}$, yielding a relatively high optical depth.

\begin{figure}
	\centering
	\includegraphics[width=\columnwidth]{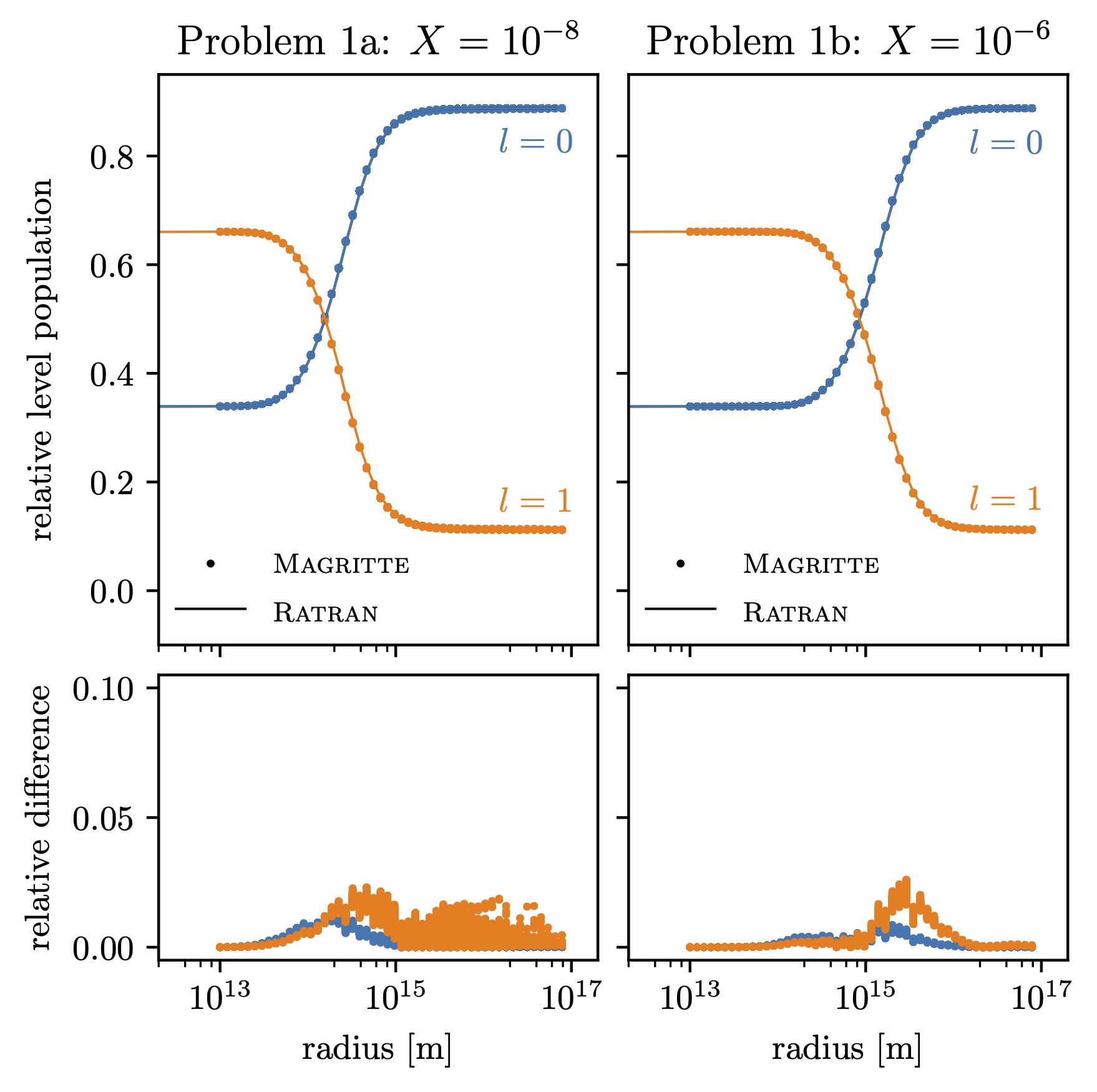}
  \caption{Comparison of the results for Problem 1 a/b of the \protect\cite{vanZadelhoff2002} benchmark obtained with \Magritte{} (dots) and \textsc{Ratran} (lines). The relative difference of two values is measured as twice the absolute difference over their sum.}
  \label{fig:VanZadelhoff_p1}
\end{figure}

Figure \ref{fig:VanZadelhoff_p1} shows a comparison between the resulting level populations for Problem 1a/b obtained with \Magritte{} and \textsc{Ratran}. Both are clearly in good agreement.

\subsubsection{Van Zadelhoff Problem 1 c/d}
\label{subsubsec:VZ_p1cd}

Since line radiative transfer models critically depend on a proper sampling of the velocity field along the line of sight of each ray, it is worthwile to test if this is properly accounted for. Therefore, we consider again benchmark problem 1 a/b from the previous paragraph, but this time with a non-zero velocity field. Although this test was not part of the \cite{vanZadelhoff2002} benchmark, we can still compare our results with \textsc{Ratran}. We consider a velocity field that is pointing radially outward, given by
\begin{equation}
    \textbf{v} (r) \ = \ \varv_{\infty} \left(\frac{r-R_{\text{in}}}{R-R_{\text{in}}}\right)^{\gamma} \hat{\textbf{r}}.
\label{eq:velfield}
\end{equation}
In the benchmarks below we used $\gamma=0.5$ and since it is the same model setup as in Section \ref{subsubsec:density_distribution} and \ref{subsubsec:VZ_p1ab} the inner radius is $R_{\text{in}} = 1.0 \times 10^{13}$ m. Furthermore, we consider two different terminal velocities $v_{\infty} =$ 10 km s$^{-1}$ and 50 km s$^{-1}$.

\begin{figure}
	\centering
	\includegraphics[width=\columnwidth]{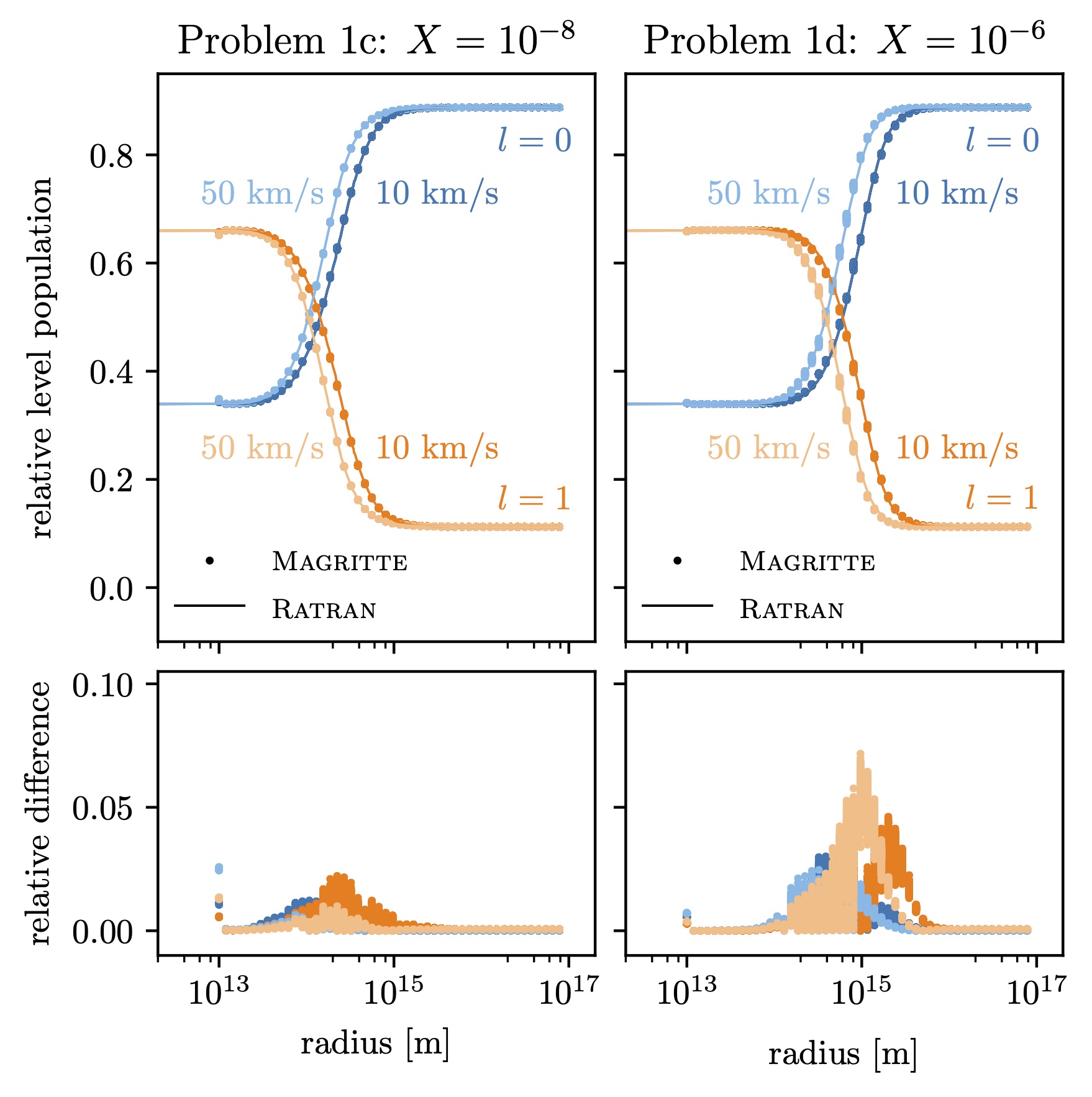}
  \caption{Comparison of the results for Problem 1 c/d obtained with \Magritte{} (dots) and \textsc{Ratran} (lines). The indicated velocities are the $\varv_{\infty}$ for each model. The relative difference of two values is measured as twice the absolute difference over their sum.}
  \label{fig:VanZadelhoff_p1_velo}
\end{figure}

Figure \ref{fig:VanZadelhoff_p1_velo} shows a comparison between the resulting level populations for Problem 1 c/d obtained with \Magritte{} and \textsc{Ratran}. Both are clearly in good agreement. However, in order to obtain this result, we needed to increase the number of grid points in the input for \textsc{Ratran} by a factor of 10 (resulting in 500 logarithmically spaced grid points). For any lower number of grid points, \textsc{Ratran} had difficulty properly sampling the velocity field and produced significantly different results from \Magritte{}.

\subsubsection{Van Zadelhoff Problem 2 a/b}
\label{subsubsec:VZ_p2ab}

The third test has a more realistic setup and considers the lines of HCO$^{+}$ in a snapshot of an inside-out collapse model by \cite{Shu1977}. This is referred to as problem 2 a/b in the benchmark paper. The parameters describing the input model were taken from the website of the benchmark\footnote{Benchmark website: \href{http://www.strw.leidenuniv.nl/astrochem/radtrans/}{www.strw.leidenuniv.nl/astrochem/radtrans/}.}. The model consists of 50 logarithmically spaced grid points. In each grid point the radial velocity, gas temperature, micro-turbulence, and HCO$^{+}$ and H$_{2}$ abundances are given. Again there are two cases, one with a relatively low optical depth where the fractional HCO$^{+}$ abundance is $X=10^{-9}$ and one with a relatively high optical depth where the relative molecular abundance is $X=10^{-8}$.

\begin{figure}
	\centering
	\includegraphics[width=\columnwidth]{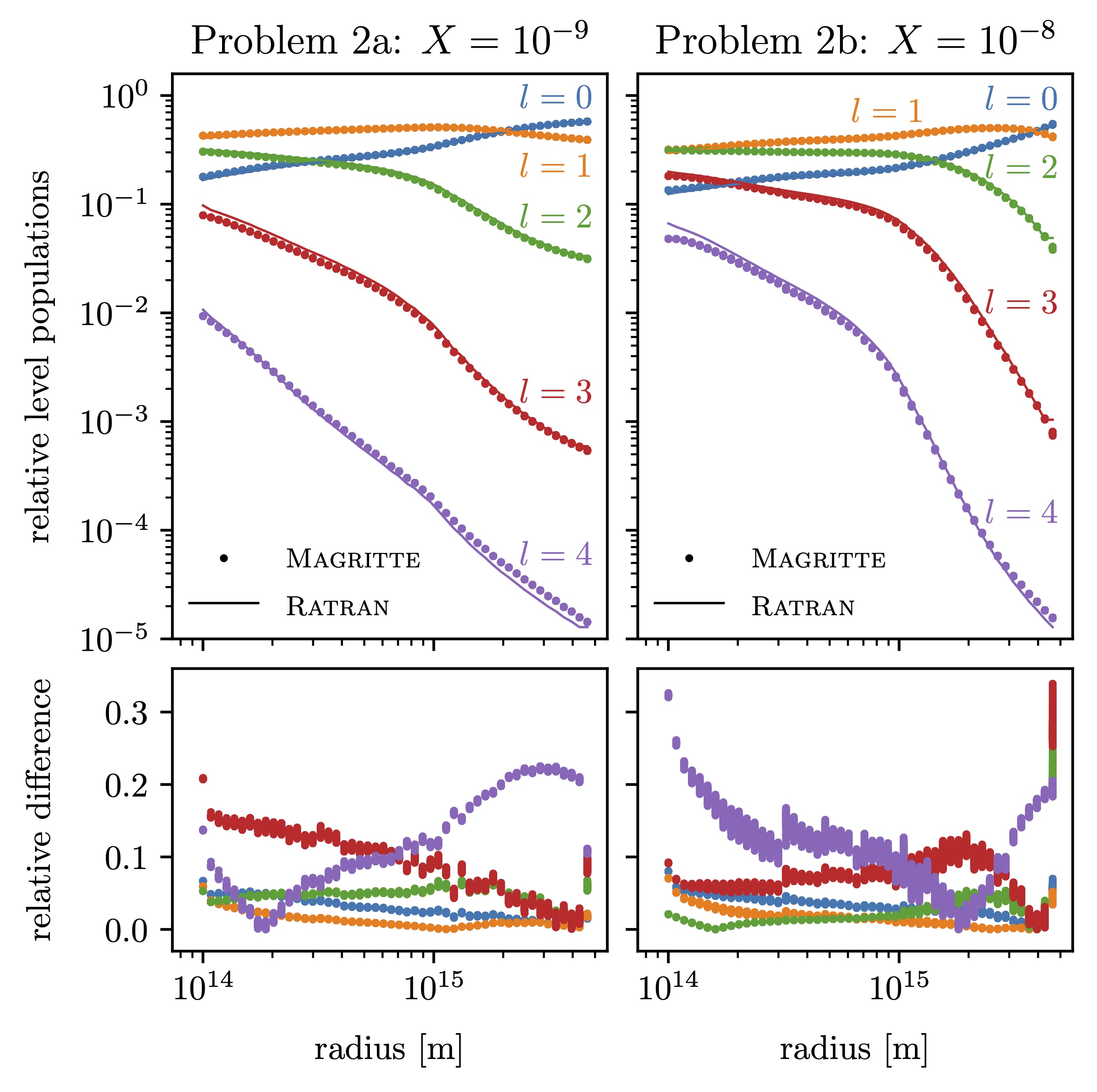}
  \caption{Comparison of the results of the first 5 levels (of 41) for Problem 2 a/b of the \protect\cite{vanZadelhoff2002} benchmark obtained with \Magritte{} (dots) and \textsc{Ratran} (lines). The relative difference of two values is measured as twice the absolute difference over their sum.}
  \label{fig:VanZadelhoff_p2}
\end{figure}

Figure \ref{fig:VanZadelhoff_p2} shows a comparison between the results for Problem 2 a/b obtained with \Magritte{} and \textsc{Ratran}. Overall, both codes agree well with relative differences below 0.3 for the first five levels. This is comparable to what \cite{Brinch2010} find in their Figure 10 for \textsc{Lime} and what \cite{Rundle2010} find in their Figures 2 and 3 for \textsc{Torus}. Furthermore, \cite{Rundle2010} report that for $l=0$ their relative deviation from the benchmark paper is less than $5\%$, which is also comparable to what we find.

\begin{table}
\centering
\caption{\Magritte{} parameters for the benchmark models.}
\label{tab:benchmark_parameters}
\begin{tabular}{l | c | c | c | c}
  \hline
  model & ($N_{\text{shells}}$) & $N_{\text{cells}}$ & $N_{\text{rays}}$ & $N_{\text{q}}$ \\ \hline
  Problem 1 a/b/c/d & 50 & 23 280 & 192 & 24 \\
  Problem 2 a/b     & 50 & 23 280 & 192 & 24
\end{tabular}
\end{table}

\section{Application}
\label{sec:application}

To demonstrate the applicability of \Magritte{} in a more realistic setup, we consider the CO line radiative transfer in a simple Keplerian disc model. This is a typical use case of 3D radiative transfer modelling \citep[see e.g.][]{Booth2019, Homan2018}  The density distribution in cylindrical ($r, \phi, z$) co-ordinates is described by
\begin{equation}
  \rho (r,\phi,z) \ =
  \begin{cases}
      0 & \ \text{for} \ \ r < r_{\text{in}}      \\
      \rho_{\text{in}} \left(\frac{r}{r_{\text{in}}}\right)^{p} \exp \left[-\frac{1}{2}\left(\frac{z}{H(r)}\right)^{2}\right] & \ \text{for} \ \ r \geq r_{\text{in}},
  \end{cases},
\end{equation}
with $p = -2.125$, and a vertical Gaussian scale height given by
\begin{equation}
  H(r) \ = \ r_{\text{in}} \ \sqrt{\frac{\kb T_{\text{in}}}{m^{\text{H}_{2}}} \frac{r_{\text{in}}}{ G M_{\star}}} \ \left( \frac{r}{r_{\text{in}}} \right)^{h},
\end{equation}
where $h=1.125$ and $m^{\text{H}_{2}}$ is the mass of H$_{2}$. The fractional CO abundance is a constant $n_{\text{CO}} = 5.0 \times 10^{-4}$. Furthermore, we assume a gas temperature distribution given by a power-law
\begin{equation}
  T(r,\phi,z) \ = \ T_{\star} \left(\frac{r}{r_{\star}}\right)^{q},
\end{equation}
in which we take $q=-0.5$, and a Keplerian velocity field
\begin{equation}
  \textbf{v}(r,\phi,z) \ = \ \sqrt{\frac{G M_{\star}}{r}} \ \bm{\hat{\phi}} .
\end{equation}
The remaining physical parameters of the star and the disc are summarised in Tables \ref{tab:star_parameters} and \ref{tab:disc_parameters} respectively.

\begin{table}
\centering
\caption{Parameters of the star in the Keplerian disc model.}
\label{tab:star_parameters}
\begin{tabular}{c c c}
  \hline
  $M_{\star}$ [$M_{\sun}$] & $T_{\star}$ [K] & $r_{\star}$ [AU] \\ \hline
  2.0                      & 2500            & 2.0
\end{tabular}
\end{table}

\begin{table}
\centering
\caption{Parameters of the Keplerian disc.}
\label{tab:disc_parameters}
\begin{tabular}{c c c}
  \hline
  $\rho_{\text{in}}$ [kg m$^{-3}$] & $T_{\text{in}}$ [K] & $r_{\text{in}}$ [AU]  \\ \hline
  $5.0\times 10^{-12}$             & 1500                & 10.0
\end{tabular}
\end{table}

Since we do not yet have a fully implemented algorithm to generate model meshes (see Section \ref{subsubsec:meshing_algorithm}), we currently use the sampling algorithm and Voronoi mesher implemented in \textsc{Lime}\footnote{Source code can be found at \href{https://github.com/lime-rt/lime}{github.com/lime-rt/lime}.} \citep{Brinch2010}.

\begin{figure*}
	\centering
	\includegraphics[width=2\columnwidth]{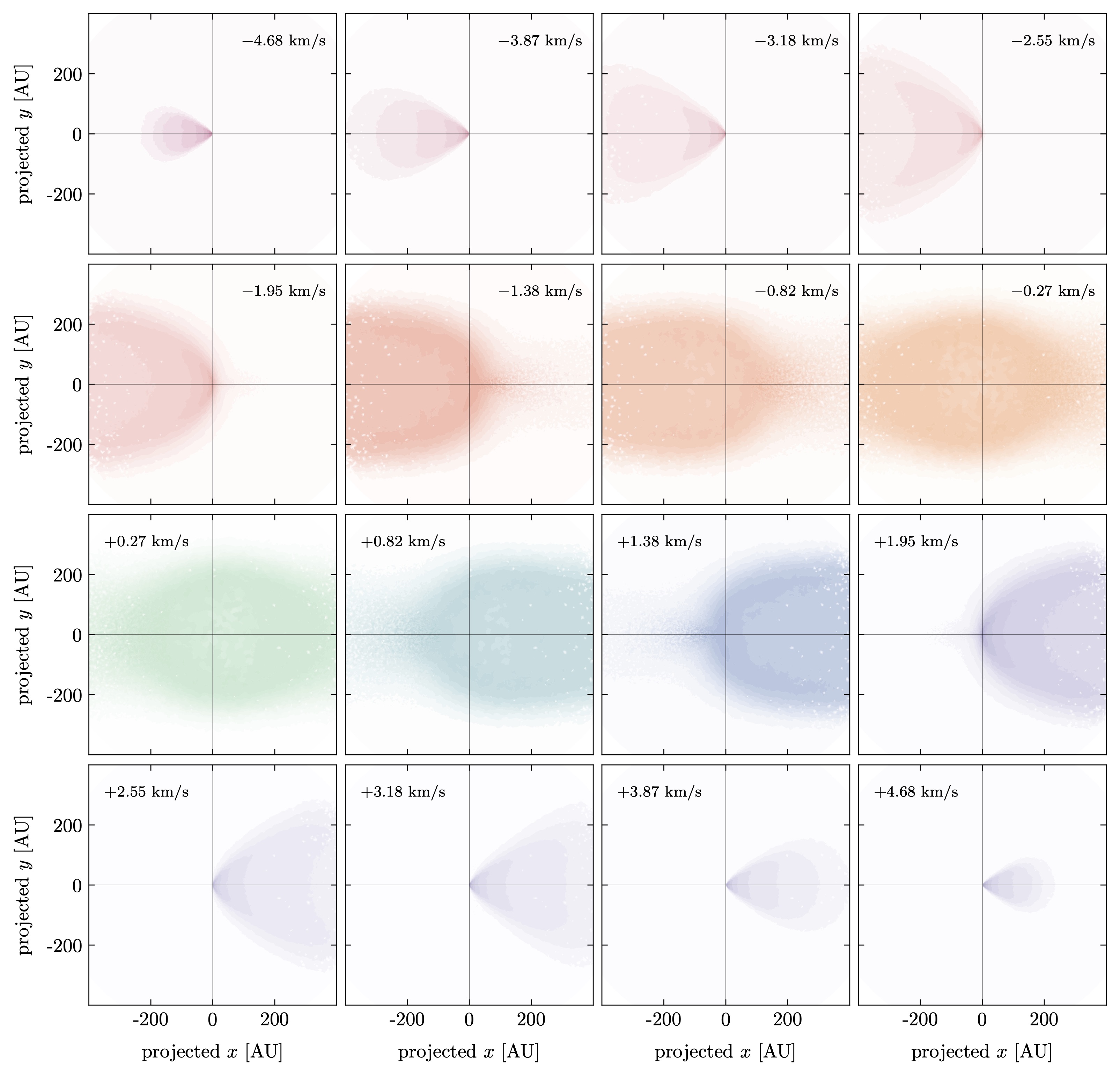}
  \caption{Channel maps with contours of the CO $J=6-5$ transition in an edge-on view of the Keplerian disc model.}
  \label{fig:disc_channel_maps}
\end{figure*}

Figure \ref{fig:disc_channel_maps} shows 16 channel maps of the CO $J=6-5$ transition in an edge-on view of the Keplerian disc model produced by \Magritte{}. From left to right and top to bottom, one can clearly see the left half of the disc moving away from the observer and being red-shifted, whereas the right half of the disc is moving towards the observer being blue-shifted.

\begin{figure}
	\centering
	\includegraphics[width=\columnwidth]{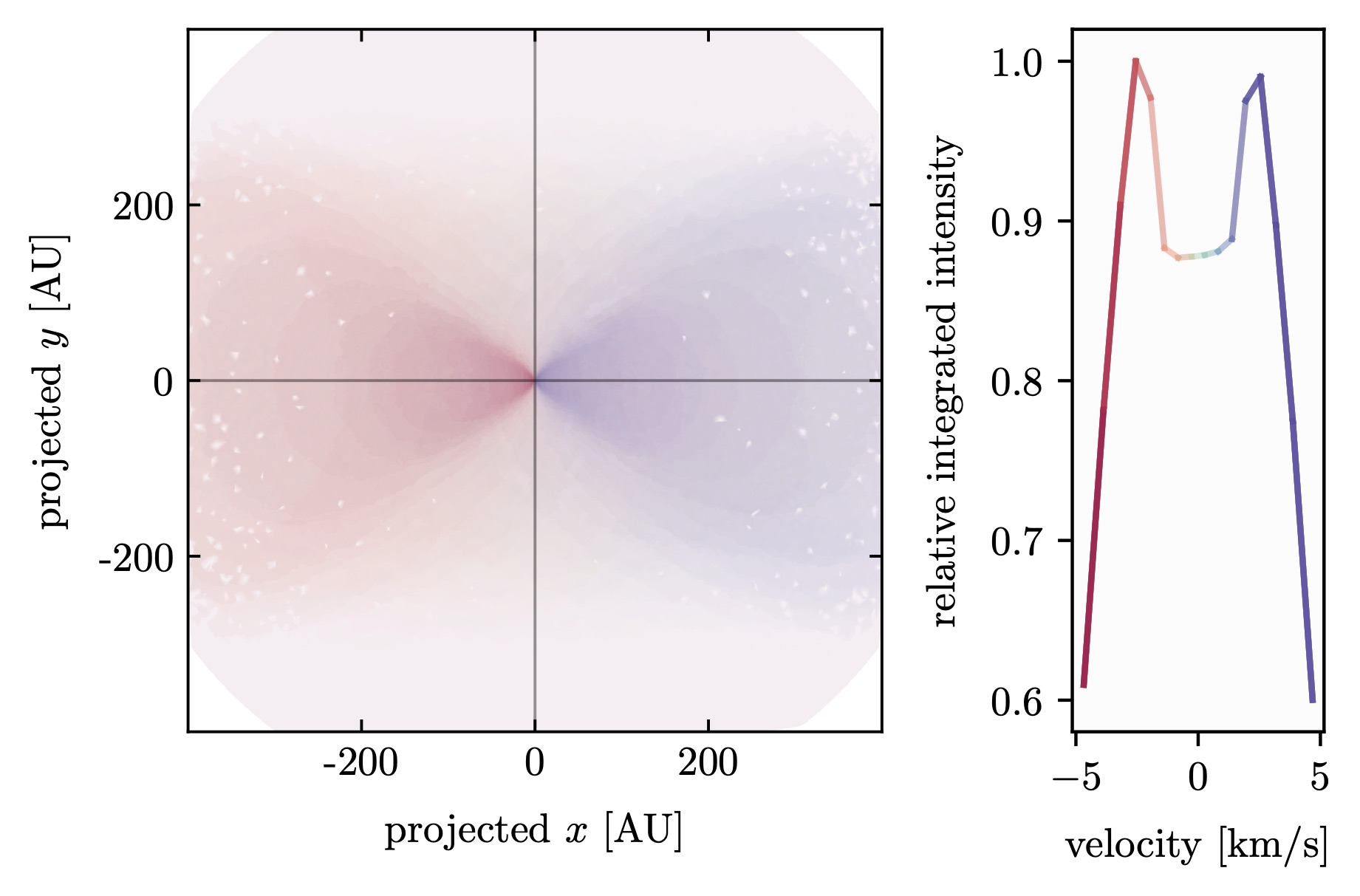}
  \caption{Composite image of the channel maps of the CO $J=6-5$ transition in the edge-on view of the Keplerian disc model \textit{(left)} and the relative integrated intensity for each of the channel maps as a function of velocity \textit{(right)}.}
  \label{fig:disc_composite}
\end{figure}

Figure \ref{fig:disc_composite} shows a composite image stacking 16 channel maps depicted in Figure \ref{fig:disc_channel_maps}, as well as the relative integrated intensity for each of the channel maps as a function of the velocity along the line of sight with respect to the rest-frame of the observer. The relative integrated intensities are normalised with respect to the maximum integrated intensity of the channel-maps.

\section{Discussion}
\label{sec:discussion}

\subsection{ALI and Ng acceleration schemes}

Both accelerated Lambda iterations (ALI) and the Ng acceleration scheme are used in \Magritte{} to ensure correct results and reduce the computation time. This is done to avoid false convergence and to reduce the number of required iterations in computing the non-LTE level populations.

In general, a wider (i.e. more non-local) band diagonal matrix ALO yields a better approximation to the Lambda operator and thus will yield better convergence.
However, when a non-local ALO is used, the resulting level populations from solving equation (\ref{eq:ourALI}) cannot be guaranteed to be positive, and thus can become unphysical \citep{Rybicki1991}.
This becomes apparent, in particular, when the solution is far from converged or when a larger bandwidth is used.
It is hard to determine in advance whether a certain bandwidth for the ALO will result in unphysical level populations.
Even the simplest two-cell model with a two-level species can easily be made to fail.
Hence, when using a non-local ALO, one should always check the validity of the level populations after solving the statistical rate equations.
If for a certain ALO, the computation yields unphysical level populations, one can always set up and solve the system of rate equations again using only a part of the ALO.
Setting up and solving the statistical rate equations only takes a fraction of the time required to compute the radiation field and the corresponding ALO.
Therefore, the trade-off that should be considered in deciding the bandwidth of the ALO is the gained reduction in iterations versus the time it costs to compute the extra off-diagonal elements.
Unfortunately, there is no generally applicable (problem independent) way to make this trade-off since the number of required iterations strongly depends on the model under consideration. \cite{Hauschildt1994} recommended optimal bandwidths for some typical model setups, however, these only apply to their particular implementation.
By default, \Magritte{} will use a diagonal (i.e. local) ALO. Larger bandwidth ALOs can be used when specifically requested. However, \Magritte{} will always check the validity of the resulting level populations and reduce the bandwidth if required. Moreover, given a model, \Magritte{} can predict how the computation time of one iteration would change when changing the bandwidth of the ALO. This should help users to decide on an appropriate bandwidth for their particular model.

The effectiveness of the Ng acceleration scheme depends on the quality of the solutions in previous iterations.
Therefore, it is advisable to start acceleration only after a certain level of convergence is already reached.
Also, the optimal balance between regular and Ng accelerated iteration steps is highly problem dependent.
It is, however, less critical than the choice of ALO bandwidth since the computational cost of taking into account more iterations in the Ng acceleration scheme is negligible (compared to the cost of computing the radiation field and the corresponding ALO).
By default, \Magritte{} will perform an Ng acceleration step after every four regular iterations, and using the level populations of all four previous iterations.

\subsection{Accuracy, precision, and re-sampling invariance}
\label{subsec:mesh_sample_invariance}

The results of the semi-analytic tests and cross-code benchmarks in Section \ref{sec:tests_and_benchmarks} clearly demonstrate the accuracy and precision of the radiative transfer and level population solver of \Magritte{}. All models were run for various numbers of cells, rays and frequency bins. Also the uniform distributions of the mesh points over the spherical shell was varied, as well as the relative distributions of the mesh points over the shells. These variations did not induce any significant differences in the results of \Magritte{}, demonstrating its re-sampling invariance. The parameters of the models presented in this paper are all about five times larger than the coarsest model that produces reasonable results.

The \cite{vanZadelhoff2002} problem 1 benchmarks with the additional velocity field, presented in Section \ref{subsubsec:VZ_p1cd}, emphasise the importance of a proper sampling of the velocity field. In \Magritte{} this is automatically taken care of by the ray-tracer which will interpolate the source and optical depth if the velocity changes too rapidly, whereas in \textsc{Ratran} the user has to provide a 10 times finer model mesh in order to obtain accurate results (comparing Figures \ref{fig:VanZadelhoff_p1_velo} and \ref{fig:VanZadelhoff_p1_velo_50gp}).
This is acceptable for a 1D solver such as \textsc{Ratran}, because the sampling of a 1D velocity field can still be assessed with comparative ease by the user. In 3D, however, velocity structures can become extremely complex along the various lines of sight. Therefore, the on-the-fly assessment and interpolation of the velocity field, as implemented in \Magritte{}, is essential to ensure accurate results.

\begin{figure}
	\centering
	\includegraphics[width=\columnwidth]{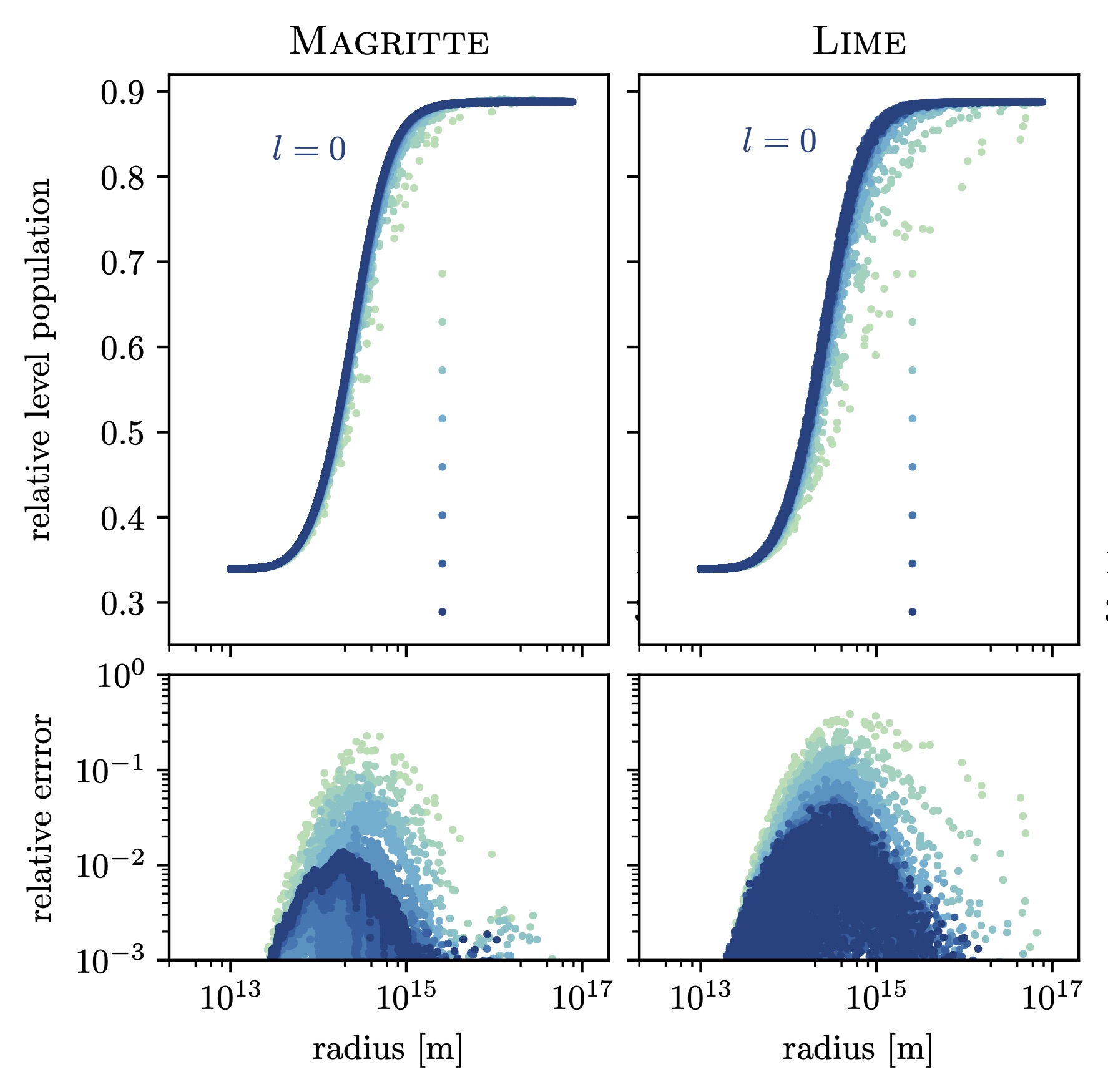}
  \caption{Comparison of the results for Problem 1 a of the \protect\cite{vanZadelhoff2002} benchmark obtained with \Magritte{} and \textsc{Lime} for different numbers of grid points. The relative error of two values is measured as twice the absolute difference with respect to the solution of \textsc{Ratran} over their sum.}
  \label{fig:VanZadelhoff_p1_sampling_Lime}
\end{figure}

Figure \ref{fig:VanZadelhoff_p1_sampling_Lime} shows a comparison between the results of \Magritte{} and \textsc{Lime}\footnote{Throughout this paper any reference to \textsc{Lime} refers to (currently latest) release version 1.9.5 (see \href{https://github.com/lime-rt/lime/releases}{github.com/lime-rt/lime/releases}).} \citep{Brinch2010} for the \cite{vanZadelhoff2002} benchmark problem 1a for different numbers of grid points. The bottom plots show the errors on the solutions, i.e. the relative difference with respect to the \textsc{Ratran} solution, which is assumed to be the most accurate. Both solvers used the exact same model mesh. In order to do this, the model was first run with \textsc{Lime} which created the mesh that could then also be used by \Magritte{}. The results for both solvers are plotted after the same number of iterations (which was around 30). The exact number was determined by the number of iterations that \Magritte{} required to reach a relative change in level populations below $10^{-7}$. In order to make a fair comparison the Ng acceleration in \Magritte{} was disabled and only a local (diagonal) ALO was used. Nevertheless, the results of \Magritte{} are clearly more accurate than the results of \textsc{Lime}. This is apparent especially for the coarser meshes with fewer mesh points. Furthermore, the results of \Magritte{} are more precise, i.e. less spread at a given radius, than the results of \textsc{Lime}. This can be attributed to the Monte Carlo noise present in the solution of \textsc{Lime}.

\begin{figure}
	\centering
	\includegraphics[width=\columnwidth]{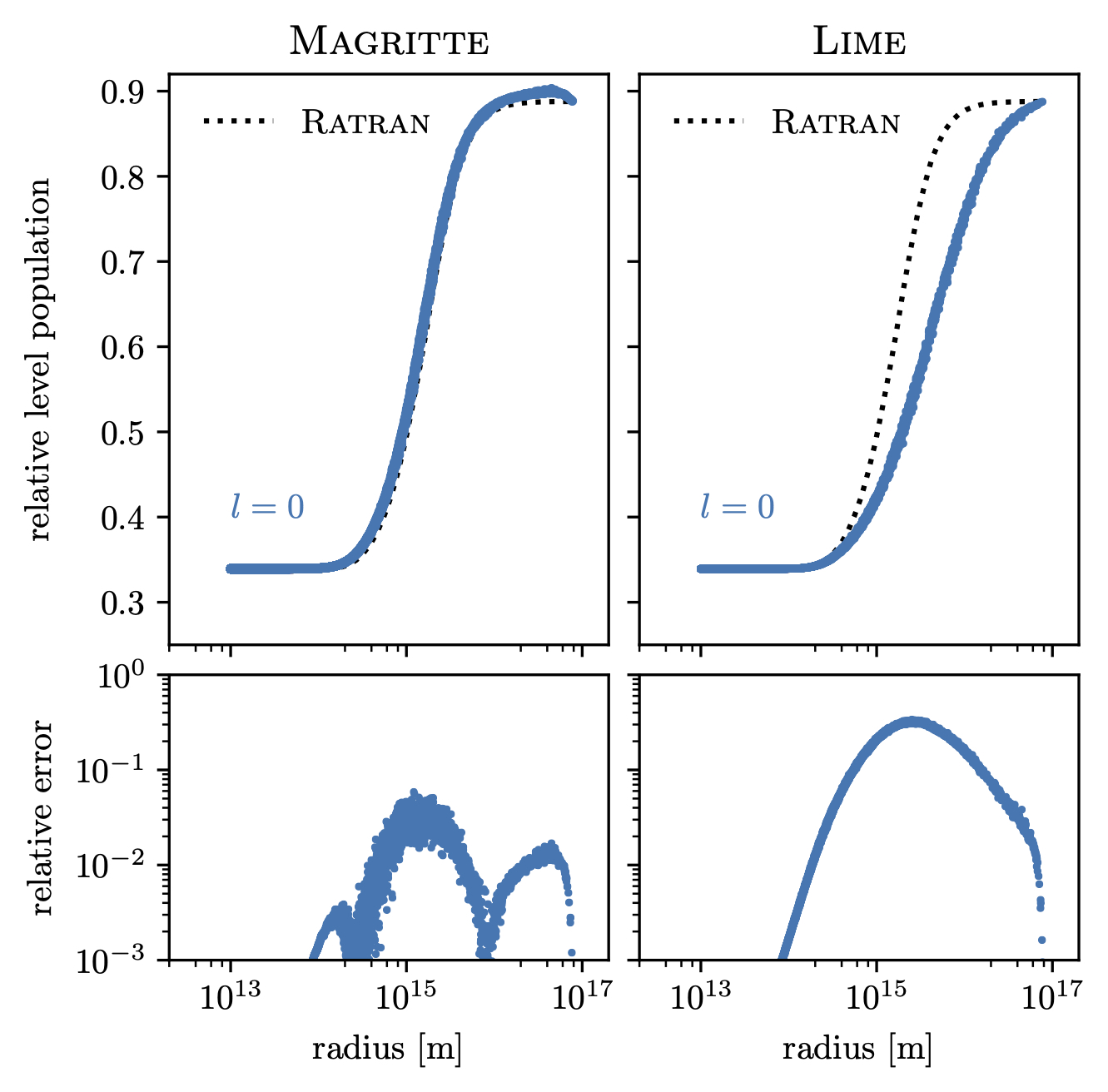}
  \caption{Comparison of the results for Problem 1 b of the \protect\cite{vanZadelhoff2002} benchmark obtained with \Magritte{} and \textsc{Lime}. The relative error of two values is measured as twice the absolute difference with respect to the solution of \textsc{Ratran} (dotted line) over their sum.}
  \label{fig:VanZadelhoff_p1b_Lime}
\end{figure}

Figure \ref{fig:VanZadelhoff_p1b_Lime} shows a comparison between the results of \Magritte{} and \textsc{Lime} for the \cite{vanZadelhoff2002} benchmark problem 1b, which is similar to problem 1a, but has a higher optical depth. Both solvers again use the same model mesh consisting of 30 000 grid points. The plot shows the results after 147 iterations (when \Magritte{} reached a relative change in level populations below $10^{-7}$). Clearly, the result of \Magritte{} is much more accurate than the result of \textsc{Lime}. This can partly be attributed to the mesh, since also the results of \Magritte{} are slightly worse than the results obtained on the mesh of shells in the comparison with \textsc{Ratran} (see Section \ref{subsubsec:VZ_p1ab}). Nevertheless, there is clear discrepancy in the accuracy that can be achieved with \Magritte{} and \textsc{Lime} in this high optical depth problem. We tried increasing the number of mesh points with a factor of 10 but did not see any improvement in the performance of \textsc{Lime}.

\subsection{Computational performance}
\label{subsec:computational_performance}

\begin{figure}
	\centering
	\includegraphics[width=.9\columnwidth]{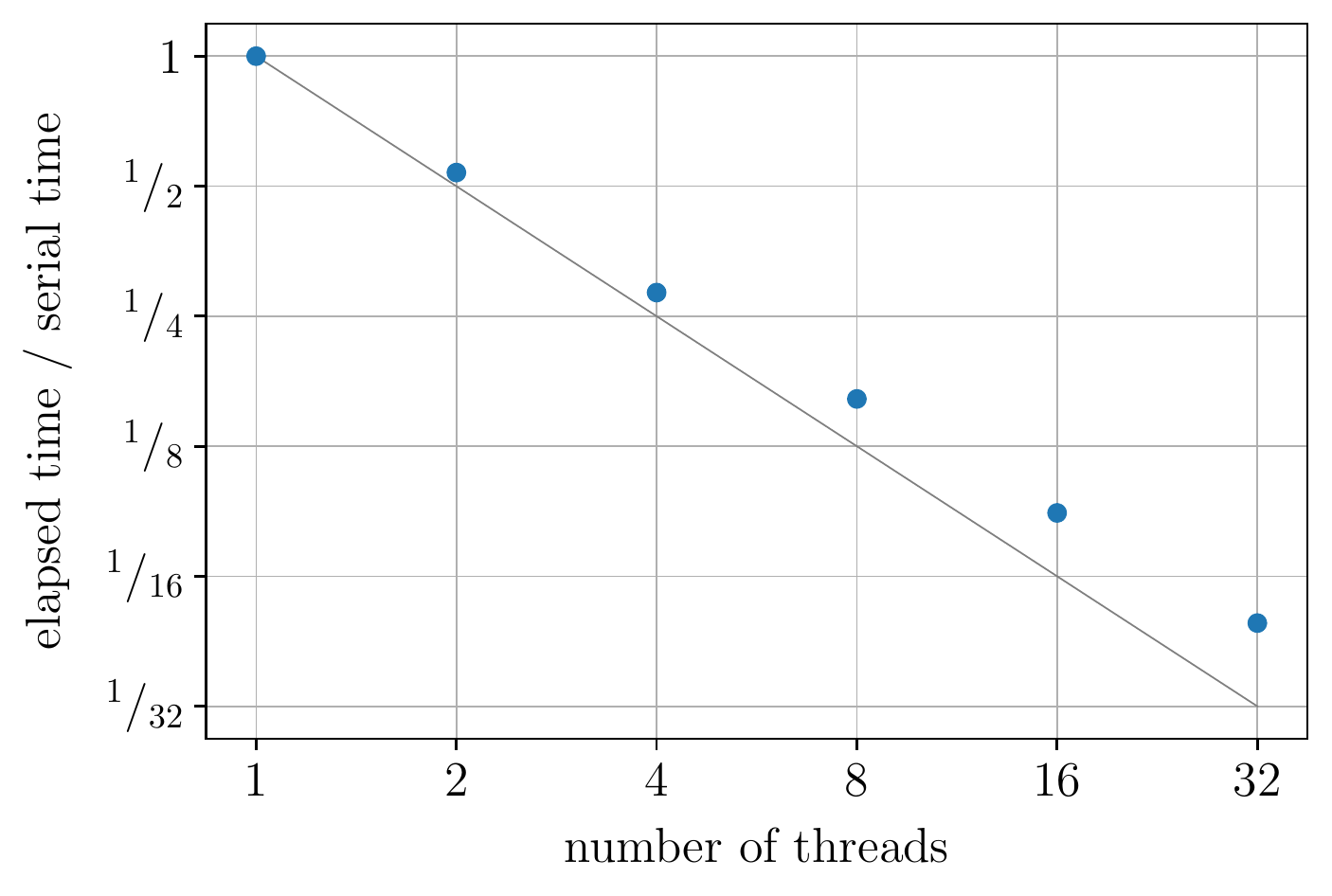}
  \caption{Plot of the (strong) scaling of \Magritte{}'s parallelisation for shared-memory systems. The dots indicate the relative timings and the grey line indicates the ideal scaling behaviour. Timings are averages over seven runs performed on one 32-core Intel Skylake node of the CSD3 cluster. (Hyper-threading was disabled for these runs such that the number of threads effectively corresponds to the number of cores used.)}
  \label{fig:strong_scaling_omp}
\end{figure}

\Magritte{} was especially designed to achieve good scalability of performance on modern distributed computer architectures and to leverage hardware acceleration.
However, since a significant part of radiative transfer research is mainly performed on commercial workstations (laptops and desktops), \Magritte{} should also perform well on these (shared memory) systems.
Figure \ref{fig:strong_scaling_omp} shows the preliminary (strong) scaling of \Magritte{} for the Keplerian disc model of Section \ref{sec:application} on a shared memory system (32-core Intel Skylake, with hyper-threading disabled).
The fact that the run time is almost perfectly inversely proportional to the number of threads shows that \Magritte{} can both effectively and efficiently use the available computational resources.
The (strong) scaling in Figure \ref{fig:strong_scaling_omp} is only preliminary in the sense that no effort was made to ensure load balancing over the cores, which could improve the scaling.
Future versions of \Magritte{} will include an active load balancing algorithm to ensure good (strong) scaling results independent of the model geometry (De Ceuster et al. \textit{in prep.}).

To gauge the computational speed of \Magritte{}, we performed and timed the \cite{vanZadelhoff2002} benchmark problem 1 (see also Section \ref{subsubsec:VZ_p1ab}) with the established 3D radiative transfer code \textsc{Lime} and compared the results with \Magritte{}.
We used the Voronoi model mesh produced by \textsc{Lime} as input for \textsc{Magritte} to ensure that both solvers got the exact same input. To further ensure a fair comparison we disabled the Ng acceleration in \Magritte{} and only used a local (diagonal) ALO. We performed 35 iterations (which in \Magritte{} corresponded to a relative change in level populations below $10^{-7}$).
We found that \Magritte{} was about 1.6 times faster than \textsc{Lime}\footnote{As measured with 8 threads on the 8 cores of a decent but standard laptop with an Intel Core i7-7700HQ CPU clocked at 2.8 GHz.} on the same mesh for the same number of iterations.
This is mainly due to the implementation with the explicit vectorisation and despite the fact the formal solver used in \Magritte{} is more precise and intrinsically slower than the one in \textsc{Lime}.
Note that we only measured the time spent in the computation of the radiation field and the level population solver and not the time spent in creating, reading or writing the model mesh.
Considering that \Magritte{} can already obtain accurate and precise results for much coarser grids (see Section \ref{subsec:mesh_sample_invariance}), we could conclude that \Magritte{} is more than 1.6 times faster. However, how much more largely depends on the required accuracy and is hard to compare between \Magritte{} and \textsc{Lime} because of the intrinsic difference in precision.

When large velocity fields are included, \Magritte{} will be slightly slower than \textsc{Lime} because of its careful treatment of the Doppler shifts along each ray. However, this careful treatment is required to obtain accurate results (see Section \ref{subsec:mesh_sample_invariance}). The current implementation heavily prioritises accuracy over speed. In future versions, the new meshing algorithm will allow us much better control over the accuracy of the radiative transfer solver, which will allow us to better balance the trade-off between accuracy and speed.

\subsection{Future development of \Magritte{}}
\label{subsec:future_development}

The current paper only reports on the first step in the development of \Magritte{}. The code base is still under active development and will be extended and improved over the next few years. The design strategy will be twofold, on the one hand focusing on developing a complete radiative transfer library with a complete modelling pipeline to confront simulations with observations, and on the other hand achieving higher performance by leveraging modern computer architectures.

\subsubsection{Meshing algorithm}
\label{subsubsec:meshing_algorithm}

The next step towards a complete radiative transfer modelling library will be to develop a mesher to generate 3D meshes for a given model (distributions for density, velocity, temperature, etc.). Since \Magritte{} is a formal solver we can build the mesh based on the numerical error it will induce in the radiative transfer computations. The algorithms used by the mesher can then also be used to assess and improve model meshes coming from hydrodynamical simulations. A strong handle on the model mesh and thus on the induced numerical errors will greatly improve the accuracy and reliability of radiative transfer computations.

\subsubsection{Including more physics}
\label{subsubsec:more_physics}

The next piece of physics to be included in \Magritte{} will be to account for scattering within the existing radiative transfer solver. Once we can account for scattering, we can do dust continuum radiative transfer and include a thermal balance module to iteratively determine the dust temperature. Later, we will focus on coupling \Magritte{} with (photo)chemistry and hydrodynamics codes to provide fully self-consistent radiation-hydro-chemical models.

\subsubsection{Computational aspects}
\label{subsubsec:computational_aspects}

The initial motivation to develop \Magritte{} was to create a general-purpose software library for 3D radiative transfer, that could leverage modern computer architectures, such as highly distributed systems with accelerators (e.g. GPUs and FPGAs), to improve the performance of astrophysical and cosmological simulations. Therefore, \Magritte{} was vectorised and parallelised for both shared and distributed memory systems, and can off-load certain computations to accelerators. The full optimisation and parallelisation strategy will be presented in a forthcoming paper. All future releases of \Magritte{} and its source code, including the optimised and accelerated versions, will be made publicly available$^{\ref{GNUgplv2}}$ at \href{https://github.com/Magritte-code}{github.com/Magritte-code}.

\section{Conclusions}
\label{sec:conclusions}

In this first paper in a series on \Magritte{}: a modern open-source software library for 3D radiative transfer modelling, we presented and tested its non-LTE line radiative transfer module. \Magritte{} uses a deterministic ray-tracer and formal solver that computes the radiation field by (iteratively) solving the radiative transfer equation along a fixed set of rays originating from each point. The ray-tracing algorithm only requires the locations of the cell centres and the nearest neighbour lists. Hence, it can readily be applied to smoothed particle hydrodynamics (SPH) particles, as well as structured and unstructured model meshes. We formulated an elegant solution method for the second-order form of the radiative transfer equation along a ray pair based on \cite{Feautrier1964} and \cite{Cannon1971, Cannon1972}, treating the scattering contributions from other rays in an iterative way. Furthermore, we presented our implementation of the accelerated Lambda iteration scheme by \cite{Rybicki1991} in this context.
We demonstrated the validity of \Magritte{} by comparing its results against both semi-analytical model solutions and the established (1D) radiative transfer solver of \textsc{Ratran} \citep{Hogerheijde2000} on the \cite{vanZadelhoff2002} benchmark for line radiative transfer. As an example application, we used \Magritte{} to generate channel maps of CO lines in a simple Keplerian disc model. Comparing our results with the established 3D radiative transfer solver \textsc{Lime} \citep{Brinch2010}, we conclude that \Magritte{} produces more accurate and more precise results, especially at high optical depth, and that it is faster.


\section*{Acknowledgements}

We are very grateful to Michiel Hogerheijde and Daniel Harsono for fruitful discussions on the line radiative transfer benchmarks and the comarison with \textsc{Ratran} and \textsc{Lime}. We also would like to thank the anonymous referee for their helpful suggestions. FDC is supported by the EPSRC iCASE studentship programme, Intel Corporation and Cray Inc. FDC, WH, and LD acknowledge support from the ERC consolidator grant 646758 AEROSOL. This work was performed using the Cambridge Service for Data Driven Discovery (CSD3), part of which is operated by the University of Cambridge Research Computing on behalf of the STFC DiRAC HPC Facility (\href{https://dirac.ac.uk}{www.dirac.ac.uk}). The DiRAC component of CSD3 was funded by BEIS capital funding via STFC capital grants ST/P002307/1 and ST/R002452/1 and STFC operations grant ST/R00689X/1. DiRAC is part of the National e-Infrastructure.



\bibliographystyle{mnras}
\bibliography{references}




\appendix
\section{Additional figure}

In this appendix we present an additional figure supporting our claims regarding the accuracy, precision and re-sample-invariance of \Magritte{} with respect to \textsc{Ratran}.

\begin{figure}
	\centering
	\includegraphics[width=\columnwidth]{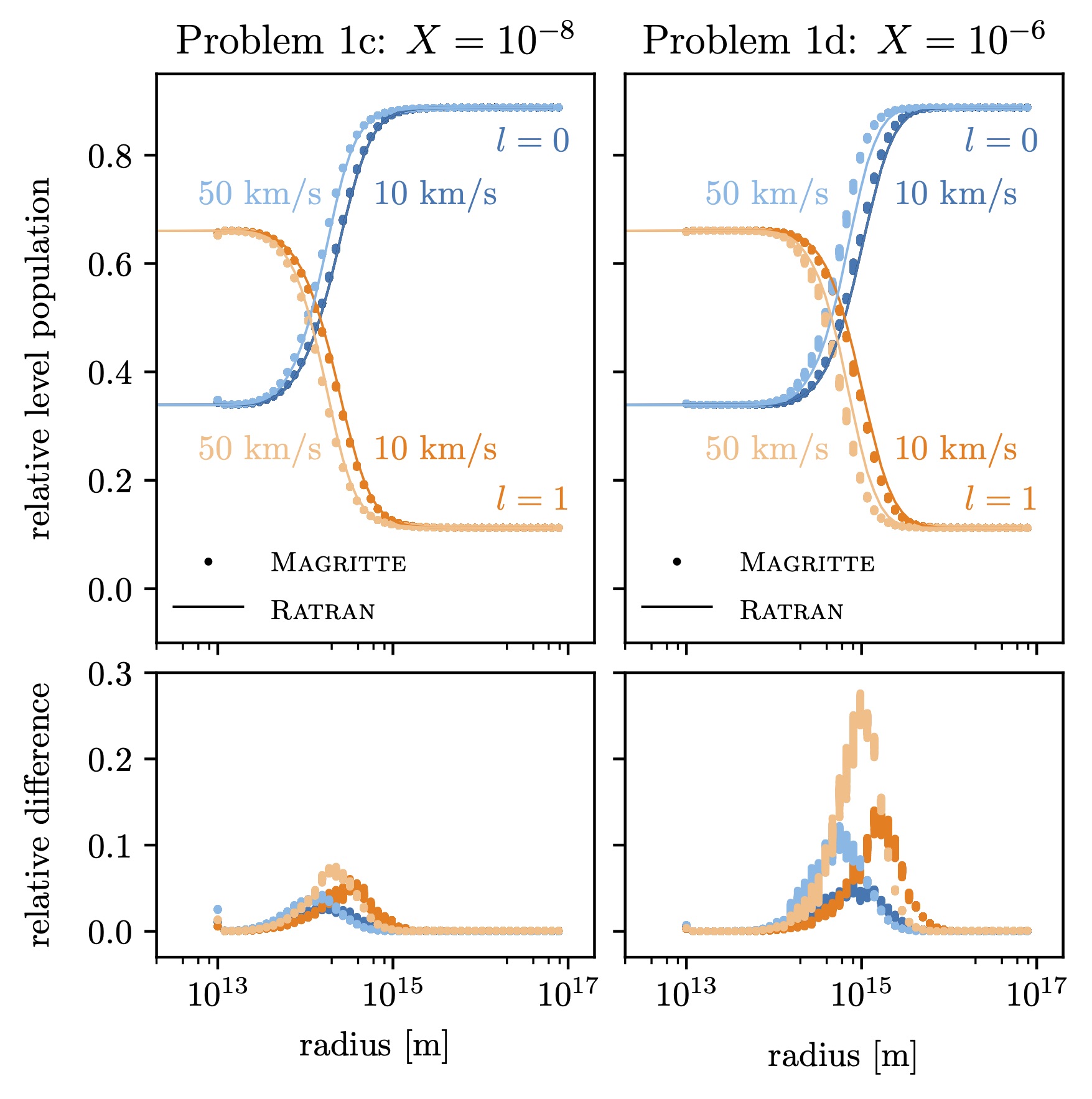}
  \caption{Comparison of the results for Problem 1 c/d, obtained with \Magritte{} (dots) and \textsc{Ratran} (lines) both on a model mesh with 50 shells. The indicated velocities are the $\varv_{\infty}$ for each model. The relative difference of two values is measured as twice the absolute difference over their sum.}
  \label{fig:VanZadelhoff_p1_velo_50gp}
\end{figure}

Figure \ref{fig:VanZadelhoff_p1_velo_50gp} shows a comparison between the results obtained with \Magritte{} (in 3D) and \textsc{Ratran} (in 1D) for a mesh with 50 shells, in contrast to figure \ref{fig:VanZadelhoff_p1_velo} in the main body of the paper where the results for a 50 shell mesh for \Magritte{} was compared to a 500 shell mesh for \textsc{Ratran}. The relative differences for the coarser \textsc{Ratran} model are about 4 times larger than for the finer model. This is due to the insufficient sampling of the velocity field which \Magritte{} can and \textsc{Ratran} cannot account for.

\bsp	
\label{lastpage}
\end{document}